\documentclass[a4paper,11pt]{article}
\usepackage{jcappub} 
\usepackage{hyperref}
\usepackage{cleveref}

\title{\boldmath Variance-aware model discrimination with the Sunyaev--Zel'dovich effect of the 21\,cm background (SZE-21cm)}

\author{C.~M.~Takalana}
\affiliation{South African Astronomical Observatory (SAAO),\\
Cape Town, South Africa}
\affiliation{Department of Physics, Stellenbosch University,\\
Matieland 7602, South Africa}
\affiliation{International Astronomical Union -- Office of Astronomy for Development (IAU--OAD),\\
Cape Town, South Africa}

\emailAdd{charlest@saao.ac.za}

\abstract{The sky-averaged redshifted 21\,cm signal from Cosmic Dawn is a uniquely sensitive tracer of early heating and ionisation, but it remains challenging to measure directly. The Sunyaev--Zel'dovich effect of the 21\,cm background (SZE--21cm) provides a complementary route: Comptonisation of the incident low-frequency background by hot electrons in galaxy clusters produces a spectral distortion that can be recovered as a difference between the line of sight through a galaxy cluster and a nearby blank-sky reference, and is therefore naturally compatible with interferometric observations. We use semi-numerical simulations of the global 21\,cm background, together with a relativistic scattering kernel built from the Maxwell--J\"uttner electron distribution, to assess how well the SZE--21cm separates physically distinct Cosmic Dawn scenarios. The model suite varies star-formation efficiency, X-ray spectral hardness, heating timing, and the suppression of low-mass sources; SZE--21cm spectra are computed over 45--200\,MHz. Separability is quantified through feature-based summaries, standardised residual spectra, and a band-integrated pairwise separability index that combines seed-to-seed scatter with an astrophysical variance layer derived from coeval ON-aperture / OFF-annulus statistics. The resulting indices are intrinsic and instrument-free, and therefore upper bounds on what a real experiment with finite thermal noise can achieve. Heating-timing variations produce the strongest separations, while the X-ray hardness variations explored here are nearly degenerate with the fiducial case once the ON--OFF variance is included. We also propagate an EDGES-like benchmark curve as a morphological stress test of the framework, not as a physically validated model, illustrating how anomalous global-signal morphologies would map into the SZE--21cm.}

\begin{document}
\maketitle
\flushbottom

\section{Introduction}

The redshifted 21\,cm line of neutral hydrogen has long been recognised as a direct probe of the thermal and ionisation history of the intergalactic medium (IGM) during the Dark Ages, Cosmic Dawn, and the Epoch of Reionisation (EoR). The principal challenge is that the cosmological signal is buried beneath astrophysical foregrounds that are orders of magnitude brighter and must be removed with extreme precision. As a result, both global-signal and interferometric measurements are highly sensitive to instrumental systematics, calibration assumptions, and residual foreground contamination.

Recent years have seen substantial progress in pushing observational limits deeper into the Cosmic Dawn regime. Interferometric experiments have reported increasingly stringent upper limits across a wide range of redshifts, including new constraints extending to $z \gtrsim 15$. The New Extension in Nan\c{c}ay Upgrading LOFAR (NenuFAR) collaboration has presented deep power-spectrum upper limits at $z\sim 17$--20, enabled by improved low-frequency calibration and foreground mitigation \cite{Munshi2025}. The Hydrogen Epoch of Reionization Array (HERA) Phase~II has begun delivering first results over an expanded 50--250\,MHz bandwidth, spanning both the EoR and Cosmic Dawn, while highlighting new classes of instrumental and analysis-driven systematics that arise as sensitivity improves \cite{Abdurashidova2025}. Complementary multi-redshift constraints from the LOw-Frequency ARray (LOFAR) continue to refine the EoR parameter space through reprocessed datasets and improved pipelines \cite{Mertens2025}. In parallel, end-to-end simulations for next-generation facilities such as the Square Kilometre Array Low-frequency telescope (SKA-Low) have matured significantly: realistic SKA-Low simulations now incorporate foreground sky models, ionospheric effects, beam chromaticity, and calibration errors within unified frameworks designed to stress-test foreground-mitigation and parameter-recovery pipelines \cite{Bonaldi2026}. Despite these advances, current results indicate that progress toward a Cosmic Dawn detection is increasingly limited by residual systematics and astrophysical variance rather than raw thermal noise. At the level of sky-averaged measurements, the interpretation of claimed global-signal detections also remains actively debated; recent analyses have revisited the physical plausibility of excess radio backgrounds, reinforcing the need for independent and complementary probes that do not rely on absolute monopole measurements \cite{Cang2025}.

A complementary observational route exploits the fact that the cosmic radio background at tens to hundreds of MHz is not a perfect blackbody: it includes a frequency-dependent 21\,cm contribution in addition to the CMB. When this background passes through the hot intracluster medium (ICM) of a galaxy cluster, it is Comptonised, producing a cluster-localised spectral distortion --- the Sunyaev--Zel'dovich effect \cite{SunyaevZeldovich1970, SunyaevZeldovich1972} of the 21\,cm background (SZE--21cm), see Colafrancesco et al.~\cite{Colafrancesco2016}. The effect is recovered as a difference between the line of sight through a galaxy cluster and a nearby blank-sky reference, which makes it naturally compatible with interferometric observations and largely insensitive to absolute calibration of the sky monopole.

The SZE--21cm has been developed theoretically and explored as an observational probe in earlier studies. The foundational treatment of the SZE--21cm and its application to reionisation-era backgrounds established the relativistic formalism and emphasised the distinctive spectral structure compared to the standard thermal SZE. Colafrancesco et al.~\cite{Colafrancesco2016} demonstrated how features in the global 21\,cm background imprint into the cluster-induced distortion. More recently, Takalana et al.~\cite{Takalana2020, Takalana2021} developed an analytic and simulation-driven framework and investigated differential detectability with low-frequency facilities, including the impact of instrumental noise and additional variance terms.

The present paper is a follow-up that asks which physically distinct Cosmic Dawn scenarios the SZE--21cm can actually separate. Rather than focusing on a single fiducial background, we run a suite of \textsc{21cmFAST} simulations that vary star-formation efficiency, X-ray spectral hardness, heating timing, and the suppression of low-mass sources, and we quantify which of these variations remain distinguishable in the SZE--21cm observable. In contrast to earlier work, we add a second variance layer derived from coeval simulation cubes (three-dimensional snapshots of the 21\,cm field at a fixed redshift), estimated from an ON-aperture / OFF-annulus statistic. This provides a conservative and observationally motivated assessment of intrinsic 21\,cm fluctuations.

This paper is organised as follows. In Section~\ref{sec:formalism} we summarise the theoretical background and mathematical formalism for the 21\,cm transition signal and its associated SZE--21cm spectral distortion, and we define the discrimination metrics used throughout, including the variance-aware pairwise separability index used to construct the pairwise distinguishability heatmap. In Section~\ref{sec:pipeline} we describe the simulation and analysis pipeline, including the model suite, the generation of global signal and SZE--21cm products, and the construction of a second variance layer from coeval cubes using an ON-aperture / OFF-annulus statistic. In Section~\ref{sec:results} we present the global 21\,cm signals, the corresponding SZE--21cm spectra, feature-based summaries, standardised residual diagnostics, the pairwise distinguishability heatmap, and the coeval-derived variance products (including robustness and stacking-oriented summaries). In Section~\ref{sec:discussion}, we interpret the model separations and degeneracies, translate the variance-aware separability indices into stacking-limited observational significance, and position the SZE--21cm alongside other Cosmic Dawn probes. We summarise our findings and conclude in Section~\ref{sec:conclusions}. 

The semi-numerical global-signal and coeval-cube products used in this work were generated with \textsc{21cmFAST} (via its Python interface \textsc{py21cmfast}) \cite{Mesinger2011,Murray2020}, and the numerical setup is summarised in Section~\ref{sec:pipeline}. We assume a flat $\Lambda$CDM cosmology throughout, as implemented in \textsc{21cmFAST}. For the simulations analysed here we adopt
$h\equiv H_0/100\,\mathrm{km\,s^{-1}\,Mpc^{-1}}=0.68$, $\sigma_8=0.82$, $\Omega_m=0.315$, $\Omega_b=0.049$, and $n_s=0.97$, with all remaining cosmological parameters set to the code defaults.

For convenience, the principal symbols used throughout the paper are: $\delta T_b$ (global 21\,cm differential brightness temperature), $\Delta T_{\rm SZE-21cm}$ (SZE--21cm spectral distortion in brightness-temperature units), $\nu^{\rm glb}_{\min}$ and $\nu^{\rm sze}_{\min}$ (frequencies of the global-signal and SZE--21cm minima), $\sigma_{\rm seed}$ (seed-to-seed scatter from \textsc{21cmFAST} initial-condition realisations), $\sigma_{\rm astro}$ (additional astrophysical variance from coeval ON-aperture / OFF-annulus statistics), $\tau_0$ (cluster Thomson optical depth) and $kT_e$ (cluster electron temperature, with $k$ the Boltzmann constant). $N_{\rm seeds}$ denotes the number of independent \textsc{21cmFAST} realisations per model, and ``seed'' refers to the random seed of the \textsc{21cmFAST} initial-condition generator. The kernel variable $s = \ln(\nu/\nu')$ encodes the logarithmic frequency shift in a single Compton scattering, with $\nu$ the observed frequency and $\nu'$ the pre-scattering frequency.


\section{SZE--21cm formalism and discrimination metrics}
\label{sec:formalism}
\subsection{The global 21\,cm transition signal}
\label{sec:21cmformalism}

The sky-averaged (global) 21\,cm signal is commonly expressed as a differential
brightness temperature relative to the CMB,
\begin{equation}
\delta T_b(z) \simeq 27 \, x_{\rm HI}(z)
\left(1 + \delta_b \right)
\left( \frac{1+z}{10} \frac{0.15}{\Omega_m h^2} \right)^{1/2}
\left( \frac{\Omega_b h^2}{0.023} \right)
\left(1 - \frac{T_\gamma(z)}{T_S(z)} \right)\,{\rm mK},
\end{equation}
where $x_{\rm HI}$ is the neutral hydrogen fraction, $\delta_b$ the baryon overdensity,
$T_\gamma$ the CMB temperature, and $T_S$ the hydrogen spin temperature
\cite{Furlanetto2006,Pritchard2012}.

The spin temperature is determined by the competition between radiative coupling
to the CMB, collisional coupling, and Ly$\alpha$ coupling via the Wouthuysen--Field effect,
\begin{equation}
T_S^{-1} = \frac{T_\gamma^{-1} + x_c T_K^{-1} + x_\alpha T_K^{-1}}
{1 + x_c + x_\alpha},
\end{equation}
where $T_K$ is the kinetic temperature of the gas, and $x_c$ and $x_\alpha$ are the collisional and Ly$\alpha$ coupling coefficients, respectively
\cite{Wouthuysen1952,Field1958}. Throughout this work, we express the signal as a function of observing frequency, using the standard mapping $1+z = \nu_{21}/\nu$, with $\nu_{21}=1420.4$\, MHz.

In this paper, we do not attempt to re-derive the full astrophysical evolution
of $\delta T_b(z)$. Instead, physically motivated realisations of the global
21\,cm signal are generated using semi-numerical simulations that self-consistently
model Ly$\alpha$ coupling, X-ray heating, and ionisation. These realisations serve as the incident backgrounds for the SZE--21cm calculations described below.

\subsection{Relativistic Comptonisation kernel}
\label{sec:kernel}

For a cluster with electron temperature $kT_e$ and Thomson optical depth $\tau$, the Comptonisation of the incident spectrum is written as a convolution with the photon redistribution kernel $P(s)$, where $s=\ln(\nu/\nu')$ is the logarithmic frequency shift produced by a single Compton scattering (with $\nu$ the scattered and $\nu'$ the incident frequency, so net upscattering corresponds to $s>0$). The scattered intensity is
\begin{equation}
I_1(\nu)=\int_{-\infty}^{+\infty} I_0(\nu\, e^{-s})\,P(s)\,\mathrm{d}s,
\label{eq:I1}
\end{equation}
with $I_0(\nu)$ the incident specific intensity \cite{Birkinshaw1999,Carlstrom2002}. The observable SZE--21cm distortion in intensity form is
\begin{equation}
\Delta I_{\rm SZE-21cm}(\nu)=\tau\,\bigl[I_1(\nu)-I_0(\nu)\bigr],
\end{equation}
which we convert to an equivalent brightness-temperature distortion $\Delta T_{\rm SZE-21cm}(\nu)$ using the Rayleigh--Jeans relation. We adopt $I_0(\nu)$ equal to the CMB blackbody plus a Rayleigh--Jeans perturbation set by the global 21\,cm signal $\delta T_b(\nu)$ (Sec.~\ref{sec:21cmformalism}); this is the same prescription used in Colafrancesco et al.~\cite{Colafrancesco2016} and Takalana et al.~\cite{Takalana2020}. The explicit form of $I_0(\nu)$ is
\begin{equation}
I_0(\nu)=\frac{2 h\nu^3}{c^2}\!\left[\exp\!\left(\frac{h\nu}{k T_{\rm CMB}}\right)-1\right]^{-1}\!+\frac{2 k\nu^2}{c^2}\,\delta T_b(\nu),
\label{eq:I0}
\end{equation}
with $T_{\rm CMB}=2.725$\,K. The CMB Planck term largely cancels in the difference $I_1-I_0$, so model-to-model differences are driven by the $\delta T_b(\nu)$ perturbation. We work throughout in the Rayleigh--Jeans regime, where the conversion between specific intensity and brightness temperature is
\begin{equation}
\Delta T(\nu) = \frac{c^2}{2 k_B \nu^2}\,\Delta I(\nu),
\label{eq:RJconv}
\end{equation}
and we use this consistently to express all SZE--21cm spectra in temperature units.

\begin{figure}
\centering
\includegraphics[width=\columnwidth]{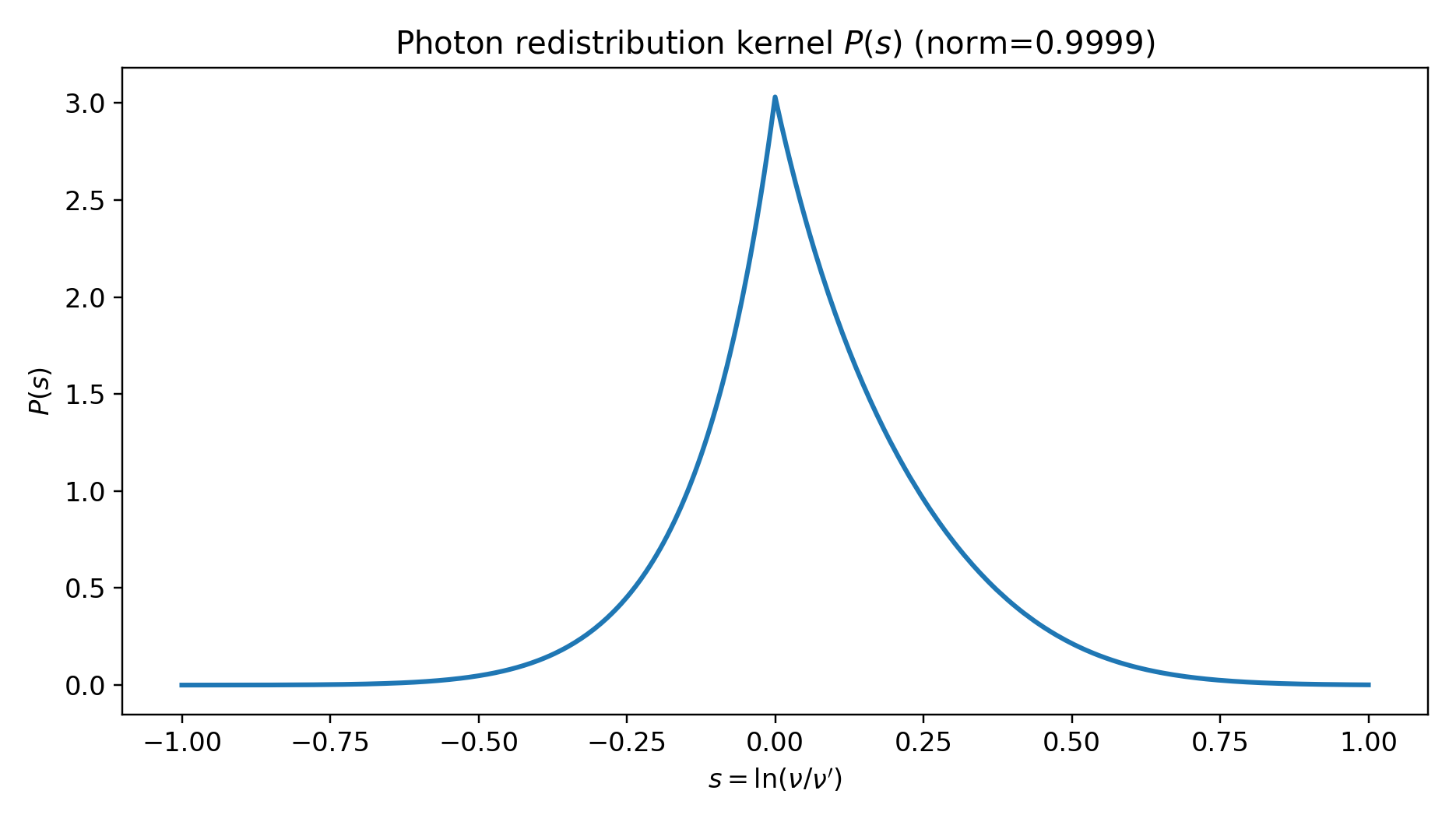}
\caption{Photon redistribution kernel $P(s)$ used to compute the relativistic SZE--21cm convolution. The kernel is normalised to $\int P(s)\,ds \simeq 0.999905$ for the adopted resolution, indicating numerically stable integration.}
\label{fig:kernel}
\end{figure}

The kernel $P(s)$ is computed from the Maxwell--J\"uttner momentum distribution of relativistic thermal electrons, following the formulation of Ensslin \& Kaiser~\cite{EnsslinKaiser2000} (see also \cite{Birkinshaw1999,Colafrancesco2003}). Defining $g \equiv m_e c^2 / kT_e$ and the dimensionless electron momentum $p$, the single-electron, single-scattering redistribution function $r(s,p)$ is
\begin{equation}
\begin{aligned}
r(s,p)=\frac{e^s}{8 p^5}\Biggl\{
&\,\frac{-3\,|1-e^s|\bigl[1+(10+8p^2+4p^4)e^s+e^{2s}\bigr]}{4 p\,e^s} \\
&+ 3(1+e^s)\!\left[\frac{3+3p^2+p^4}{\sqrt{1+p^2}}-\frac{(3+2p^2)\bigl(2\ln(p+\sqrt{1+p^2})-|s|\bigr)}{2 p}\right]\!\Biggr\},
\end{aligned}
\label{eq:rsp}
\end{equation}
defined for $|s| \le 2\ln(p+\sqrt{1+p^2})$ and zero otherwise. The kernel is then the thermal average over the Maxwell--J\"uttner distribution,
\begin{equation}
P(s)=\frac{\int_0^\infty p^2\,e^{-g\sqrt{1+p^2}}\,r(s,p)\,\mathrm{d}p}{\int_0^\infty p^2\,e^{-g\sqrt{1+p^2}}\,\mathrm{d}p}.
\label{eq:Ps}
\end{equation}
Equation~(\ref{eq:Ps}) is evaluated numerically on a uniform $(s,p)$ grid with Simpson-rule quadrature. The resulting kernel for $kT_e=10$\,keV is shown in Fig.~\ref{fig:kernel}: the achieved normalisation is $\int P(s)\,\mathrm{d}s = 0.999905$, consistent with stable numerical integration and adequate for the convolution that follows.

We have additionally checked the normalisation and the first two moments of $P(s)$ against the non-relativistic limit of the Ensslin \& Kaiser~\cite{EnsslinKaiser2000} kernel, $\langle s\rangle = 3\theta_e$ and $\langle s^2\rangle = 2\theta_e$ with $\theta_e = kT_e/m_e c^2$, finding agreement to better than $1\%$ on the first moment at $kT_e=10$\,keV and a deviation in the second moment at the $\sim$10\% level consistent with the known $\mathcal{O}(\theta_e^2)$ relativistic correction. The convention used here normalises $P(s)$ as a probability density on $s=\ln(\nu/\nu')$ and differs from the Comptonisation kernel of Sarkar, Chluba \& Lee~\cite{SarkarChlubaLee2019}, in which a Doppler factor is absorbed into the definition; in their convention the leading Kompaneets coefficients are recovered as $\langle s\rangle_K=4\theta_e$ and $\langle s^2\rangle_K = 2\theta_e$. Both conventions yield the same observable SZE distortion when used with the convolution prescription appropriate to each. As an end-to-end test, our SZE--21cm spectra reproduce the canonical result of Colafrancesco et al.~\cite{Colafrancesco2016} in the $\delta T_b\to 0$ limit.

\subsection{Discrimination metrics}

\paragraph{Feature summary.}
We characterise each model with a small set of interpretable spectral features:
\begin{itemize}
\item $\nu^{\rm glb}_{\min}$ and $\delta T_{b,\min}$: the frequency and value of the minimum in the global 21\,cm signal.
\item $\nu^{\rm sze}_{\min}$ and $\Delta T_{\min}$: the frequency and value of the minimum in the SZE--21cm spectrum.
\item Integrated absolute ``area'' metrics over the band, e.g.\ $\int|\delta T_b|\,d\nu$ and $\int|\Delta T_{\rm SZE-21cm}|\,d\nu$.
\end{itemize}

\paragraph{Standardised residuals.}
For each model $m$, we compute a standardised residual relative to the fiducial model (\texttt{lcdm\_fiducial}, hereafter denoted by the subscript ``fid''):
\begin{equation}
Z_m(\nu)=\frac{\mu_m(\nu)-\mu_{\rm fid}(\nu)}{\sqrt{\sigma_m^2(\nu)+\sigma_{\rm fid}^2(\nu)}},
\end{equation}
where $\mu(\nu)$ denotes the mean spectrum over seeds and $\sigma(\nu)\equiv\sigma_{\rm seed}(\nu)$ the seed-to-seed scatter from independent \textsc{21cmFAST} initial-condition realisations.

\paragraph{Pairwise distinguishability heatmap.}
We define a pairwise band-integrated separability index between any two models $i$ and $j$:
\begin{equation}
D_{ij}^2=\sum_{\nu\in{\rm grid}}\frac{\left[\mu_i(\nu)-\mu_j(\nu)\right]^2}{\sigma_i^2(\nu)+\sigma_j^2(\nu)+\sigma_{\rm astro}^2(\nu)},
\label{eq:Dij}
\end{equation}
and visualize $D_{ij}$ as a matrix heatmap. The additional variance term $\sigma_{\rm astro}(\nu)$ is derived from coeval cubes (Section~\ref{sec:variance}). This produces a conservative, physically motivated assessment of separability in the presence of intrinsic 21\,cm fluctuations within ON/OFF regions.

The sum in Eq.~(\ref{eq:Dij}) is evaluated on a uniform frequency grid over 45--200\,MHz. Because the grid spacing is constant across models, the discrete sum is proportional to the band-integrated distance and we omit an explicit $\Delta\nu$ factor. We do not interpret $D_{ij}$ as a formal detection significance, but as a relative, variance-weighted separability metric suitable for ranking model distinguishability under an explicit astrophysical variance budget.

\section{Simulation and analysis pipeline}
\label{sec:pipeline}
Unless otherwise stated, we compute SZE--21cm spectra for a fixed representative cluster across the model suite, so that differences between models reflect only the incident 21\,cm background. We adopt $\tau_0 = 10^{-3}$ (a value typical of massive clusters and consistent with the working choice in Colafrancesco et al.~\cite{Colafrancesco2016} and Takalana et al.~\cite{Takalana2020}; cluster-to-cluster variations in $\tau_0$ enter as an overall multiplicative scaling of $\Delta T_{\rm SZE-21cm}$ and do not affect inter-model separability) and an electron temperature $kT_e = 10$\,keV, representative of hot, X-ray-luminous clusters such as Coma. The temperature enters the relativistic kernel of Sec.~\ref{sec:kernel} through $g=m_e c^2/kT_e$: higher $kT_e$ broadens $P(s)$ and increases the asymmetry between up- and down-scattering, while lower $kT_e$ recovers the thermal Rayleigh--Jeans limit. Cluster-population scatter in $\tau_0$ and $kT_e$ would broaden any stacked SZE--21cm signal but, in our framework, would not alter the relative ranking of model pairs.

\subsection{\textsc{21cmFAST} setup and numerical choices}
\label{sec:21cmfastsetup}

We generate the global 21\,cm signals and the coeval-cube products (three-dimensional brightness-temperature fields evaluated at a single redshift) using the semi-numerical code \textsc{21cmFAST} v3 \cite{Mesinger2011,Park2019} through the Python wrapper \textsc{py21cmfast} \cite{Murray2020}. The astrophysical parametrisation is the eight-parameter atomic-cooling galaxy model introduced by Park et al.~\cite{Park2019}; we vary the four parameters listed in Table~\ref{tab:astroparams} ($f_{\star,10}$, $L_X$, $\alpha_X$, $M_{\rm turn}$) one at a time relative to the fiducial set, holding all other parameters at their v3 defaults. We do not enable the molecular-cooling-galaxy framework of Davies et al.~\cite{Davies2025} that is now available in \textsc{21cmFAST} v4; including a separate Pop\,III population would broaden the predicted family of low-frequency global signals and is left to future work. The seven \textsc{21cmFAST}-based models therefore probe a deliberately conservative, well-controlled portion of the atomic-cooling parameter space.

All models are run on the same numerical grid to ensure that differences are driven by astrophysical assumptions rather than resolution effects. Our baseline configuration uses a comoving box size of $L_{\rm box}=300\,\mathrm{cMpc}$ with an initial density grid of $200^3$ cells and an ionisation grid of $100^3$ cells, with initial conditions controlled by the \textsc{21cmFAST} random seed. For the ON/OFF variance layer, we compute coeval cubes at the redshifts corresponding to the observing grid, and apply the same aperture/annulus geometry across models (Section~\ref{sec:variance}).

\begin{table}[!htbp]
\caption{Astrophysical parameter values used in each model. Values not listed are kept at the \textsc{21cmFAST} v3 defaults of Park et al.~\cite{Park2019} (i.e.\ $\log_{10}\alpha_\star = 0.5$, $\log_{10} f_{\rm esc,10}=-1.0$, $\log_{10}\alpha_{\rm esc} = -0.5$, $t_\star=0.5$). The ionising efficiency \texttt{HII\_EFF\_FACTOR} is fixed at $30$ for all \textsc{21cmFAST} models. Notation: $f_{\star,10}$ is the stellar-to-halo mass fraction normalised at $M_h = 10^{10}\,M_\odot$; $L_X$ is the soft-band X-ray luminosity per star-formation rate (units $\mathrm{erg\,s^{-1}\,M_\odot^{-1}\,yr}$); $\alpha_X$ is the X-ray spectral energy index; $M_{\rm turn}$ is the halo-mass turnover scale below which star formation is suppressed.}
\label{tab:astroparams}
\centering
\begin{tabular}{lcccc}
\hline\hline
Model & $\log_{10} f_{\star,10}$ & $\log_{10} L_X$ & $\alpha_X$ & $\log_{10}(M_{\rm turn}/M_\odot)$ \\
\hline
\texttt{lcdm\_fiducial}       & $-1.5$ & $40.0$ & $1.5$ & $8.7^\dagger$ \\
\texttt{high\_SFE}            & $-1.0$ & $40.0$ & $1.5$ & $8.7^\dagger$ \\
\texttt{low\_SFE}             & $-2.2$ & $40.0$ & $1.5$ & $8.7^\dagger$ \\
\texttt{very\_early\_heating} & $-1.5$ & $42.0$ & $1.5$ & $8.7^\dagger$ \\
\texttt{very\_late\_heating}  & $-1.5$ & $38.0$ & $1.5$ & $8.7^\dagger$ \\
\texttt{soft\_Xray}           & $-1.5$ & $40.0$ & $2.0$ & $8.7^\dagger$ \\
\texttt{hard\_Xray}           & $-1.5$ & $40.0$ & $1.0$ & $8.7^\dagger$ \\
\texttt{suppressed\_sources}  & $-2.0$ & $40.0$ & $1.5$ & $10.5$ \\
\hline
\end{tabular}
\\[2pt]
\raggedright {\footnotesize $^\dagger$ \textsc{21cmFAST} v3 default.}
\end{table}

\begin{table}[!htbp]
\caption{Summary of the model suite and the corresponding SZE--21cm minimum frequencies and depths. All \textsc{21cmFAST} models share $N_{\rm seeds}=6$ initial-condition realisations and the same kernel normalisation $\int P(s)\,\mathrm{d}s = 0.999905$; the EDGES benchmark is a single deterministic curve ($N_{\rm seeds}=1$). The ``Family'' column groups together models that vary the same physical parameter; the parameter values are given in Table~\ref{tab:astroparams}. The $\Delta T_{\min}$ and $\nu(\Delta T_{\min})$ columns summarise results that are discussed in Section~\ref{sec:results}.}
\label{tab:modelsummary}
\centering
\begin{tabular}{llcc}
\hline\hline
Model & Family & $\Delta T_{\min}$ (mK) & $\nu(\Delta T_{\min})$ (MHz) \\
\hline
\texttt{lcdm\_fiducial}       & fiducial   & $-0.1260$ & $64.44$ \\
\texttt{high\_SFE}            & SFE        & $-0.1277$ & $60.55$ \\
\texttt{low\_SFE}             & SFE        & $-0.1215$ & $71.24$ \\
\texttt{very\_early\_heating} & heating    & $-0.1130$ & $92.13$ \\
\texttt{very\_late\_heating}  & heating    & $-0.1370$ & $66.87$ \\
\texttt{soft\_Xray}           & X-ray      & $-0.1257$ & $64.44$ \\
\texttt{hard\_Xray}           & X-ray      & $-0.1264$ & $64.44$ \\
\texttt{suppressed\_sources}  & sources    & $-0.1231$ & $69.29$ \\
\texttt{EDGES\_benchmark}     & EDGES-like & $-0.2257$ & $63.46$ \\
\hline
\end{tabular}
\end{table}

\subsection{Model suite}

We simulate eight physically motivated Cosmic Dawn scenarios plus an EDGES-like benchmark curve (model labels in typewriter font match the figure legends and Tables~\ref{tab:modelsummary}--\ref{tab:features}):
\begin{enumerate} 
\item \texttt{lcdm\_fiducial}: $\Lambda$CDM-like baseline with the Park et al.~\cite{Park2019} 21cmFAST default astrophysical parameters listed in Table~\ref{tab:astroparams}.
\item \texttt{high\_SFE} / \texttt{low\_SFE}: increased / decreased star-formation efficiency, $\log_{10} f_{\star,10} = -1.0$ and $-2.2$ respectively (fiducial $-1.5$); modifies overall emissivity and the timing of Ly$\alpha$ coupling and heating.
\item \texttt{soft\_Xray} / \texttt{hard\_Xray}: X-ray spectral index $\alpha_X = 2.0$ (soft) and $1.0$ (hard), against the fiducial $\alpha_X = 1.5$; varies the mean free path of heating photons.
\item \texttt{very\_early\_heating} / \texttt{very\_late\_heating}: $\log_{10}(L_X/\mathrm{erg\,s^{-1}\,M_\odot^{-1}\,yr})=42$ and $38$ respectively (fiducial $40$); shifts the heating transition earlier or later relative to coupling.
\item \texttt{suppressed\_sources}: $\log_{10}(M_{\rm turn}/M_\odot) = 10.5$ together with $\log_{10} f_{\star,10}=-2.0$, suppressing low-mass haloes and delaying coupling and heating.
\item \texttt{EDGES\_benchmark}: a flattened-Gaussian template after Bowman et al.~\cite{Bowman2018} (centre 78\,MHz, depth 500\,mK, width 19\,MHz, flattening parameter $\tau=7$) propagated through the SZE--21cm formalism. This curve is included purely as a morphological stress test of the separability framework and is not interpreted as a physically validated astrophysical model; the physical plausibility of an EDGES-like absorption depth has been challenged in subsequent analyses (e.g.\ \cite{Cang2025}).
\end{enumerate}

The global 21\,cm signals are generated on a common frequency grid spanning 45--200\,MHz. We then compute the corresponding SZE--21cm spectra using the relativistic Comptonisation kernel of Sec.~\ref{sec:kernel} for a representative cluster with optical depth $\tau_0$ and temperature $kT_e$ (the same across models to isolate background-driven differences). For each model we generate $N_{\rm seeds}=6$ realisations and store the per-seed $\delta T_b(\nu)$ and $\Delta T_{\rm SZE-21cm}(\nu)$ curves together with their seed-averaged means $\mu_{\delta T_b}(\nu)$, $\mu_{\rm SZE}(\nu)$ and seed-to-seed scatters $\sigma_{\delta T_b}(\nu)$, $\sigma_{\rm SZE}(\nu)$. The seed scatter in $\Delta T_{\rm SZE-21cm}$ is small for the adopted setup (typical $\langle\sigma_{\rm SZE}\rangle\sim 5\times 10^{-5}$\,mK), reflecting the smooth, global nature of the incident background in the monopole treatment.

\subsection{Second variance layer from coeval cubes}
\label{sec:variance}

A crucial limitation of pure-monopole treatments is that real cluster observations necessarily sample finite sky regions and therefore include spatial fluctuations of the 21\,cm background. To incorporate this, we compute an additional variance layer from coeval cubes:
\begin{enumerate}
\item Generate coeval cubes at the redshifts corresponding to the observing frequency grid.
\item Create 2D maps (slice or line-of-sight averaged, depending on cube type).
\item Define an ON region (aperture) centred on the cluster and an OFF region (annulus) in surrounding blank sky.
\item Compute ON/OFF statistics and their covariance: $\mathrm{Var}(\mathrm{ON})$, $\mathrm{Var}(\mathrm{OFF})$, and $\mathrm{Cov}(\mathrm{ON},\mathrm{OFF})$.
\end{enumerate}
The ON region is defined as a circular aperture of radius 6 pixels, with the OFF region an annulus spanning radii of 8--12 pixels, where pixels refer to the projected coeval map resolution (with one pixel corresponding to $L_{\rm box}/N_{\rm pix}$ in comoving units). For each redshift, 100 independent ON/OFF placements are sampled by randomly drawing aperture centres across the 2D map to estimate the intrinsic variance and covariance. These quantities are combined into a frequency-dependent ON--OFF variance contribution,
\begin{equation}
\sigma^2_{\rm astro}(\nu)=\mathrm{Var}(\mathrm{ON})+\mathrm{Var}(\mathrm{OFF})-2\,\mathrm{Cov}(\mathrm{ON},\mathrm{OFF}),
\end{equation}
which we denote $\sigma_{\rm astro}(\nu)$. Over 45--200\,MHz, $\sigma_{\rm astro}(\nu)$ ranges from $\sim0.11$\,mK at the low-frequency end to $\sim2.72$\,mK at the high-frequency end, with a median $\sim1.35$\,mK. This term enters the pairwise separability index of Eq.~(\ref{eq:Dij}) and provides a conservative variance budget against which model differences are evaluated.

The variance products are summarised visually in Section~\ref{sec:results} to show both the frequency dependence and the scale relative to the mean SZE--21cm amplitude. This is important because it sets the practical variance budget against which broadband separability should be judged, independent of the numerical seed scatter that arises within the monopole background pipeline.

\section{Results}
\label{sec:results}

\subsection{Global 21\,cm signals}

\begin{figure}
\centering
\includegraphics[width=\columnwidth]{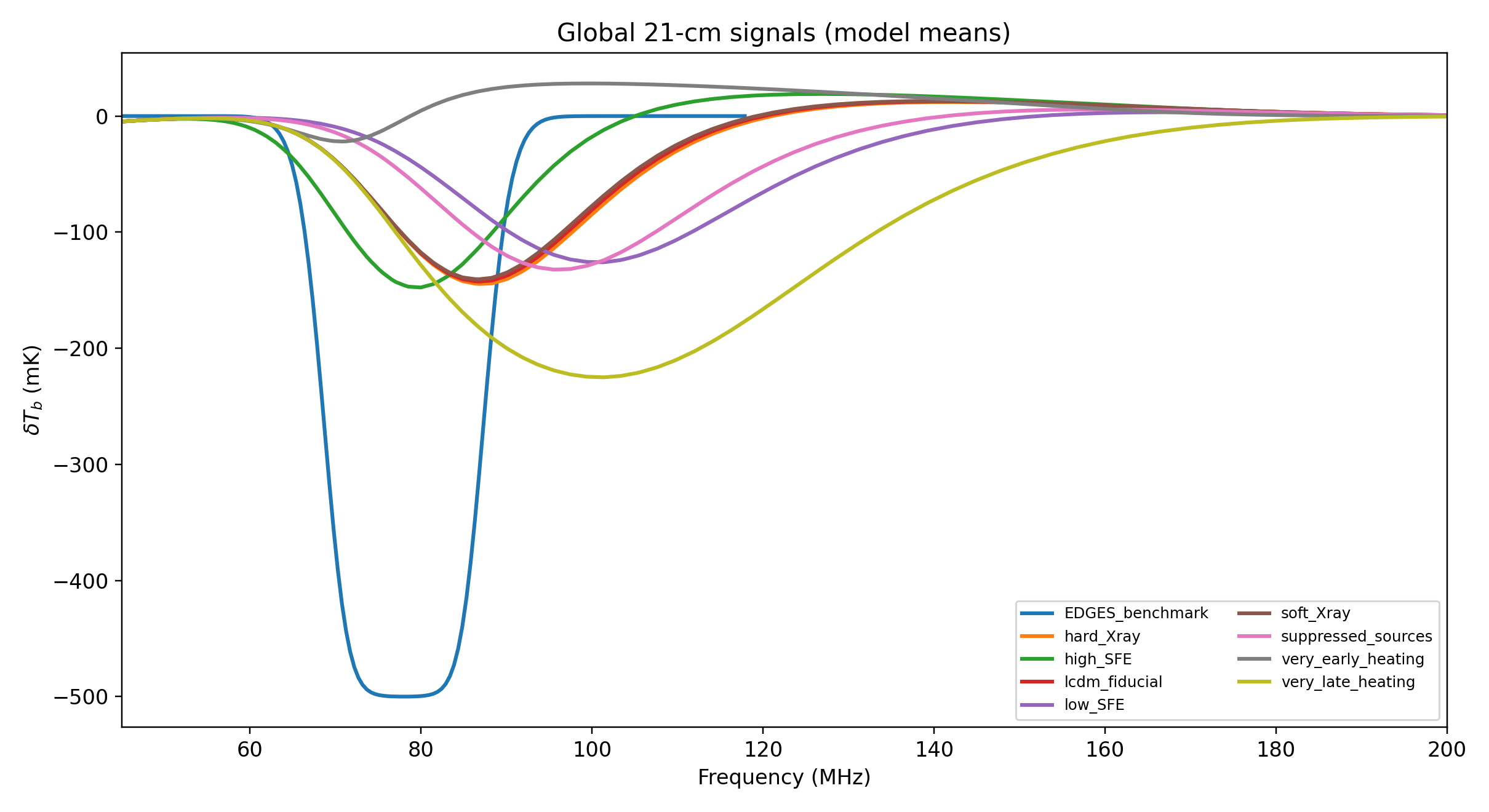}
\caption{Mean global 21\,cm signals $\mu_{\delta T_b}(\nu)$ for the full model suite over 45--200\,MHz, including the EDGES-like benchmark curve. Heating-timing variations strongly modify both the depth and the frequency location of the absorption trough.}
\label{fig:globaloverlay}
\end{figure}

Figure~\ref{fig:globaloverlay} overlays the mean global signals for all models. The fiducial model attains a minimum $\delta T_{b,\min}\approx -142.5$\,mK at $\nu \approx 86.8$\,MHz. Variations in heating timing produce large changes: the \texttt{very\_early\_heating} model yields a much shallower absorption trough ($\delta T_{b,\min}\approx -21.9$\,mK), while \texttt{very\_late\_heating} produces a deep trough ($\delta T_{b,\min}\approx -225.0$\,mK) shifted toward higher frequency ($\nu\approx101.4$\,MHz). The EDGES benchmark exhibits a deep absorption near $\nu\approx78$\,MHz by construction.

\subsection{SZE--21cm spectra}

\begin{figure}
\centering
\includegraphics[width=\columnwidth]{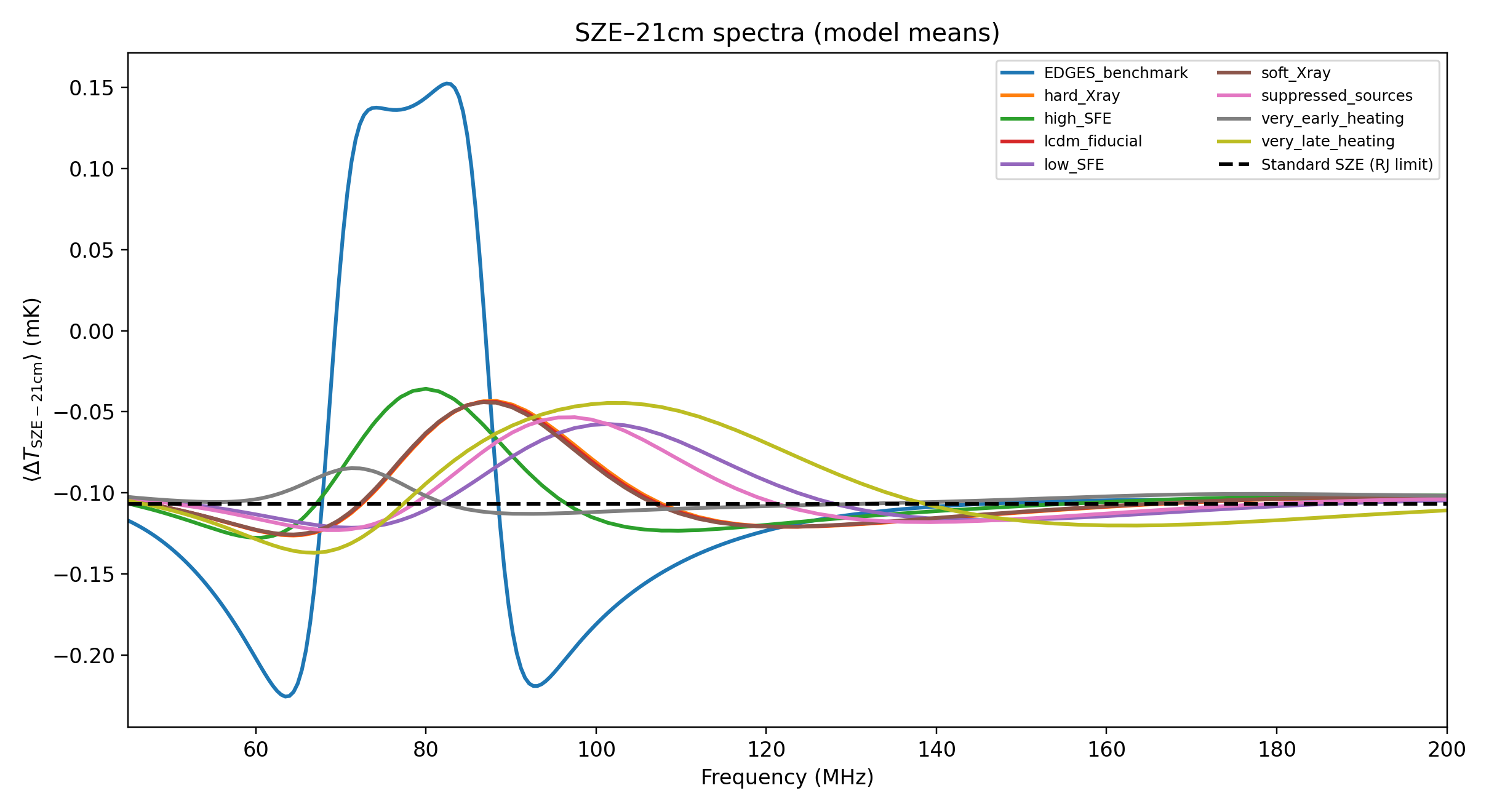}
\caption{Mean SZE--21cm spectra $\mu_{\rm SZE}(\nu)$ over 45--200\,MHz. The physically motivated models span $\Delta T_{\min}\approx-0.113$ to $-0.137$\,mK, while the EDGES benchmark yields a stronger minimum of $\approx-0.226$\,mK.}
\label{fig:szeoverlay}
\end{figure}

Figure~\ref{fig:szeoverlay} shows the corresponding SZE--21cm mean spectra $\mu_{\rm SZE}(\nu)$. Across the physically motivated model suite, the SZE--21cm minima range from $\Delta T_{\min}\approx -0.113$\,mK (\texttt{very\_early\_heating}) to $\Delta T_{\min}\approx -0.137$\,mK (\texttt{very\_late\_heating}), with the fiducial case at $\Delta T_{\min}\approx -0.126$\,mK. The EDGES benchmark yields a significantly stronger minimum $\Delta T_{\min}\approx -0.226$\,mK.

\begin{figure}
\centering
\includegraphics[width=\columnwidth]{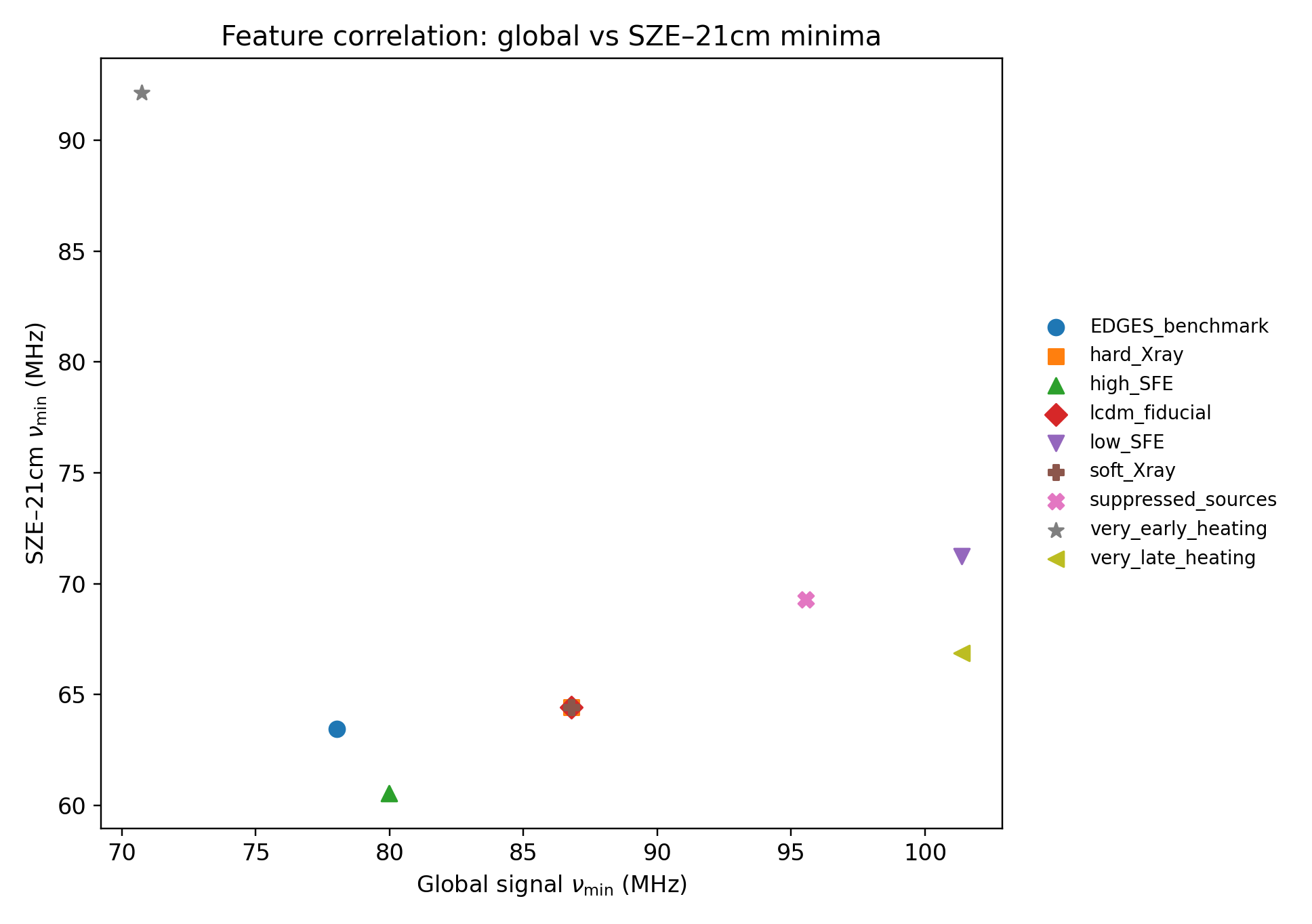}
\caption{Feature correlation between the global-signal minimum frequency $\nu^{\rm glb}_{\min}$ and the SZE--21cm minimum frequency $\nu^{\rm sze}_{\min}$ across the model suite. Deviations from a one-to-one mapping illustrate that the SZE--21cm minimum is not a direct copy of the global-signal trough position, but a convolutional response that depends on spectral curvature and the redistribution kernel.}
\label{fig:nuMinCorrelation}
\end{figure}

Table~\ref{tab:features} summarises key features for each model, including $\nu^{\rm glb}_{\min}$ and $\nu^{\rm sze}_{\min}$. Notably, the frequencies of SZE minima are systematically lower than the global-signal minima for several models (e.g.\ fiducial $\nu^{\rm sze}_{\min}\approx64.4$\,MHz vs $\nu^{\rm glb}_{\min}\approx86.8$\,MHz), reflecting the convolution nature of Comptonisation and the frequency-dependent mapping from incident spectral curvature to scattered distortions; the model-by-model relationship between the two minimum frequencies is shown in Fig.~\ref{fig:nuMinCorrelation}.

\begin{table*}
\caption{Summary of characteristic features for the global 21\,cm signal and the corresponding SZE--21cm spectra over the 45--200\,MHz band.}
\label{tab:features}
\centering
\begin{tabular}{lcccc}
\hline\hline
Model &
$\delta T_{b,\min}$ &
$\Delta T_{\min}$ &
$\int |\delta T_b|\,\mathrm{d}\nu$ &
$\int |\Delta T_{\rm SZE}|\,\mathrm{d}\nu$ \\
 &
(mK) at $\nu^{\rm glb}_{\min}$ (MHz) &
(mK) at $\nu^{\rm sze}_{\min}$ (MHz) &
(mK\,MHz) &
(mK\,MHz) \\
\hline
lcdm\_fiducial       & $-142.50$ at $86.79$ & $-0.1260$ at $64.44$ & $4663.45$  & $15.94$ \\
high\_SFE            & $-147.63$ at $79.98$ & $-0.1277$ at $60.55$ & $4463.11$  & $15.98$ \\
low\_SFE             & $-125.89$ at $101.36$& $-0.1215$ at $71.24$ & $5086.06$  & $15.83$ \\
very\_early\_heating & $-21.89$ at $70.75$  & $-0.1130$ at $92.13$ & $1937.25$  & $16.11$ \\
very\_late\_heating  & $-224.98$ at $101.36$& $-0.1370$ at $66.87$ & $12403.32$ & $15.34$ \\
soft\_Xray           & $-140.53$ at $86.79$ & $-0.1257$ at $64.44$ & $4585.35$  & $15.94$ \\
hard\_Xray           & $-144.60$ at $86.79$ & $-0.1264$ at $64.44$ & $4751.30$  & $15.93$ \\
suppressed\_sources  & $-132.20$ at $95.53$ & $-0.1231$ at $69.29$ & $4962.23$  & $15.86$ \\
EDGES\_benchmark     & $-500.00$ at $78.04$ & $-0.2257$ at $63.46$ & $9613.52$  & $19.88$ \\
\hline
\end{tabular}
\end{table*}

\subsection{Model separability in the SZE--21cm observable}

We quantify separability through difference spectra, standardised residuals, and a pairwise distinguishability heatmap.

\begin{figure}
\centering
\includegraphics[width=\columnwidth]{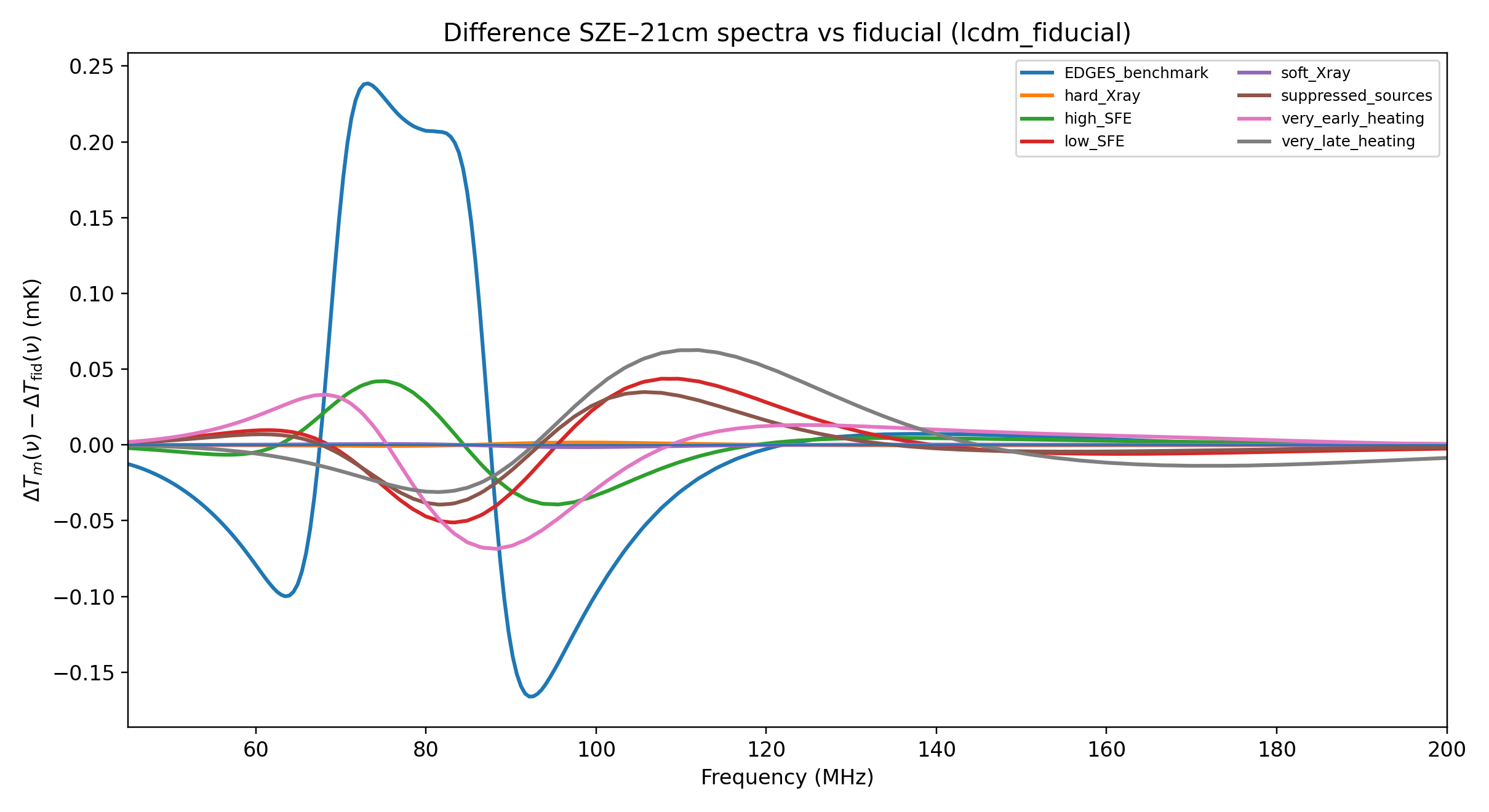}
\caption{Difference spectra relative to the fiducial case: $\mu_m(\nu)-\mu_{\rm fid}(\nu)$. Heating timing and suppressed source models produce the strongest broadband departures.}
\label{fig:diff}
\end{figure}

\begin{figure}
\centering
\includegraphics[width=\columnwidth]{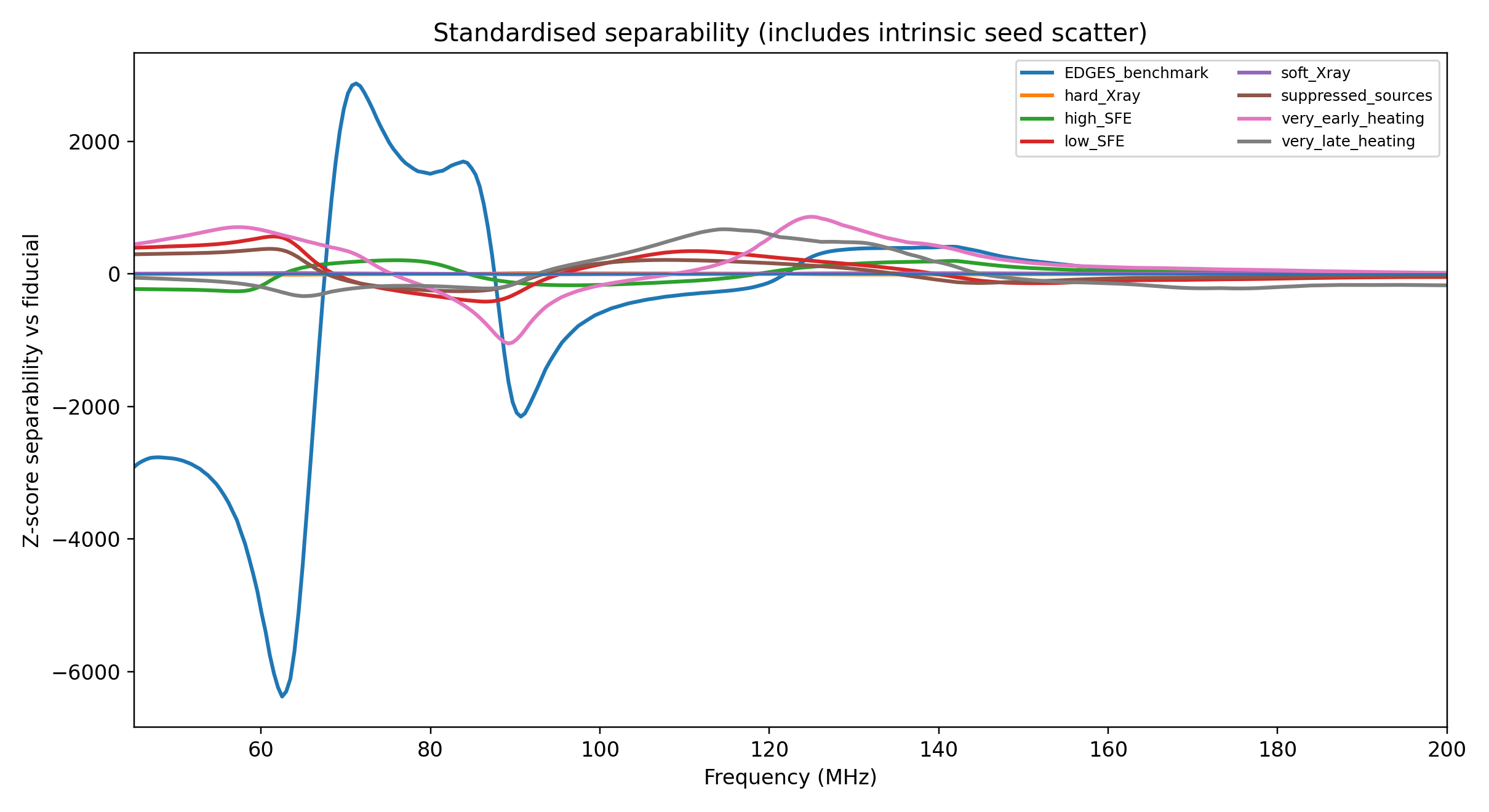}
\caption{Standardised separability versus fiducial, $Z_m(\nu)$, computed using the seed scatter in the monopole pipeline. This emphasises frequency ranges where models differ most strongly in units of intrinsic realisation scatter. In the present setup the seed scatter is small, so $Z_m(\nu)$ mainly traces mean differences; instrumental and ON/OFF variance terms can be added to the denominator in observational forecasts.}
\label{fig:zscore}
\end{figure}

\paragraph{Difference spectra and standardised residuals.}
Figure~\ref{fig:diff} shows $\mu_m(\nu)-\mu_{\rm fid}(\nu)$ for each model. Heating timing yields the most prominent spectral differences across the band. Figure~\ref{fig:zscore} shows the standardised separability $Z_m(\nu)$, isolating differences relative to seed scatter. For the current pipeline, seed scatter is tiny and $Z_m(\nu)$ primarily reflects mean differences.

\begin{figure*}
\centering
\includegraphics[width=0.90\textwidth]{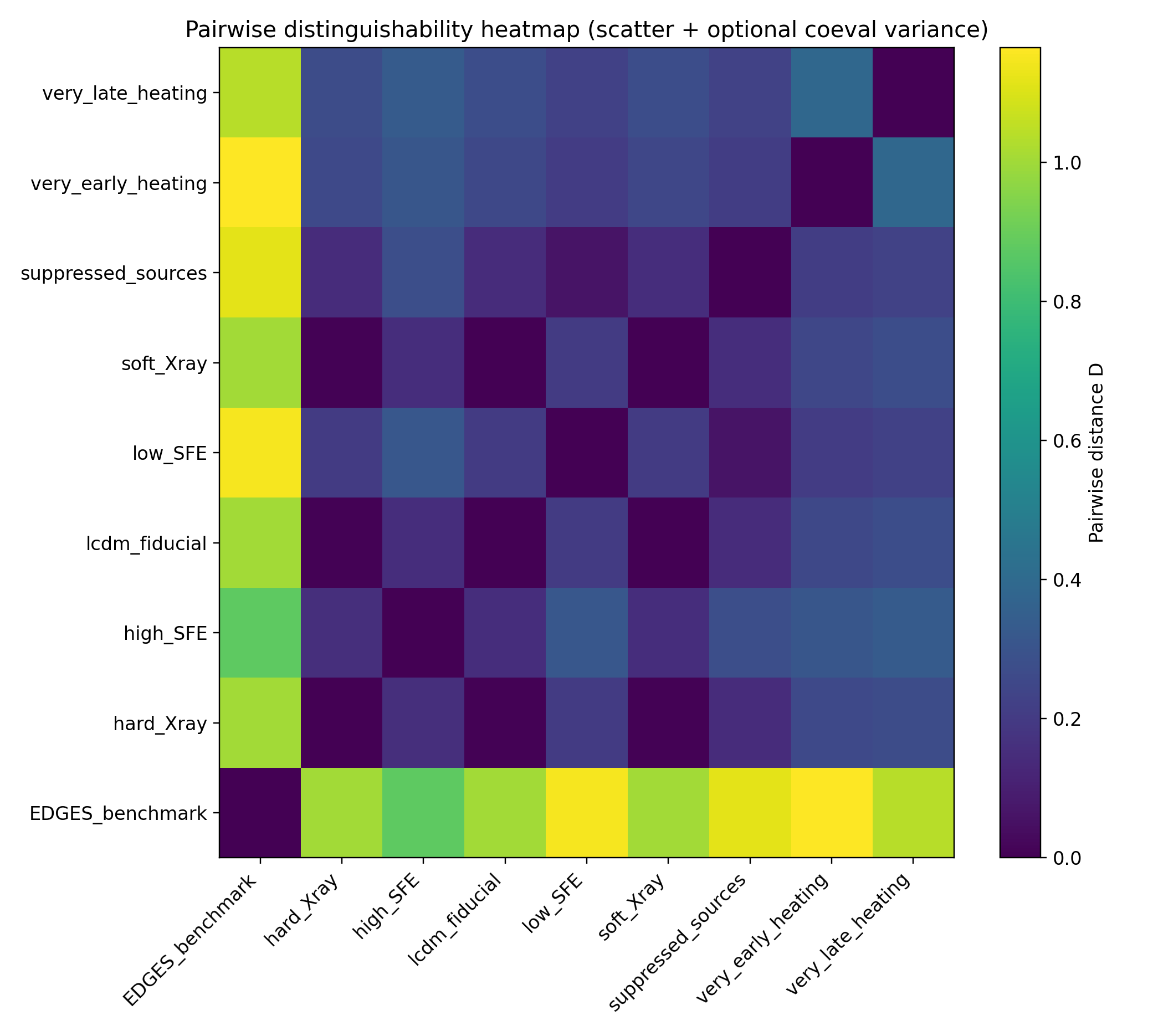}
\caption{Pairwise distinguishability heatmap for the model suite. Each element shows $D_{ij}$ from Eq.~(\ref{eq:Dij}), including seed scatter and the coeval-derived ON/OFF variance layer $\sigma_{\rm astro}(\nu)$. Heating-timing models and the EDGES benchmark are most separated from fiducial, while the particular soft/hard X-ray variants explored here remain nearly degenerate with fiducial under the adopted variance budget.}
\label{fig:heatmap}
\end{figure*}

\paragraph{Pairwise distinguishability heatmap.}
Figure~\ref{fig:heatmap} displays the pairwise distance matrix $D_{ij}$ (Eq.~\ref{eq:Dij}), computed using the seed scatter and the additional coeval-derived variance layer $\sigma_{\rm astro}(\nu)$. This heatmap provides an at-a-glance summary of which model pairs are intrinsically separable in the presence of realistic ON/OFF fluctuations.

The fiducial model is most distant from the EDGES benchmark ($D \approx 1.0$), while the soft/hard X-ray variants explored here remain nearly degenerate with the fiducial case ($D\approx 0.004$--$0.005$): the specific hardness contrast adopted does not yield a large broadband SZE--21cm separation once $\sigma_{\rm astro}$ is included. Heating-timing models are substantially separated from fiducial ($D\approx 0.25$--$0.28$), and star-formation-efficiency variations produce intermediate separations ($D\approx 0.15$--$0.20$). The \texttt{low\_SFE} and \texttt{suppressed\_sources} models are close to each other ($D\approx 0.063$), consistent with both representing reduced effective emissivity from low-mass haloes and delayed coupling and heating. We caution that the values of $D$ here are intrinsic, instrument-free separability indices and should not be read as detection significances; the translation to a stacking-limited observational separation is given in Sec.~\ref{sec:Dinterp}.

\begin{figure}
\centering
\includegraphics[width=\columnwidth]{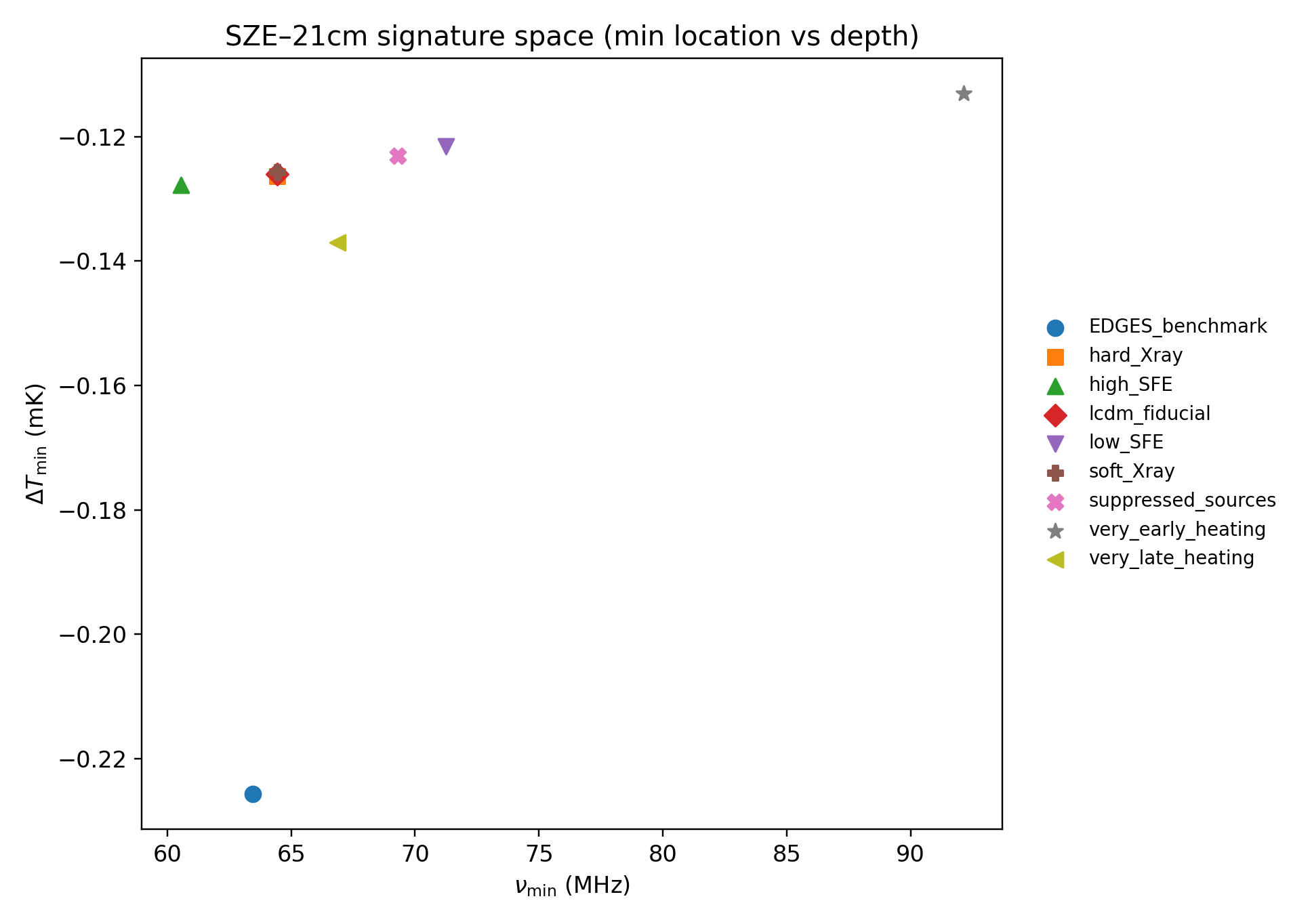}
\caption{SZE--21cm ``signature space'' showing the minimum frequency $\nu^{\rm sze}_{\min}$ versus minimum depth $\Delta T_{\min}$ for each model.}
\label{fig:signature}
\end{figure}

\paragraph{Signature space.}
Figure~\ref{fig:signature} plots the SZE--21cm minimum depth vs.\ minimum frequency as a compact ``signature space.'' Heating timing models occupy distinct loci in this plane, while X-ray hardness variants cluster close to the fiducial point for the present parameter choices.

\subsection{Second variance layer: ON--OFF coeval scatter}

\begin{figure}
\centering
\includegraphics[width=\columnwidth]{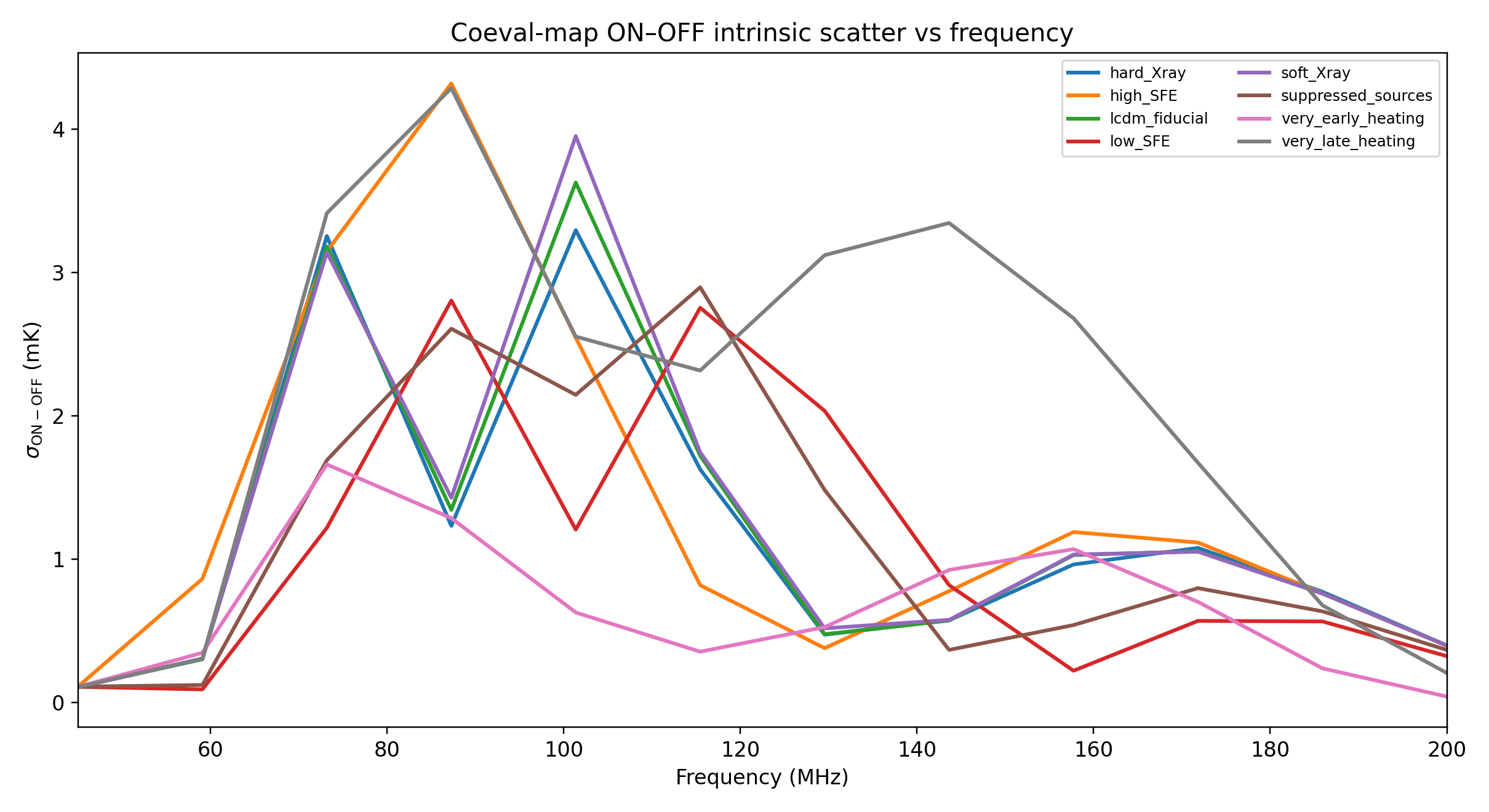}
\caption{Frequency-dependent ON--OFF intrinsic scatter derived from coeval cubes, shown here as $\sigma_{\rm ON-OFF}(\nu)$. This term captures finite-region 21\,cm fluctuations that contribute an additional variance layer beyond monopole seed scatter.}
\label{fig:coevalsigma}
\end{figure}

\begin{figure}
\centering
\includegraphics[width=\columnwidth]{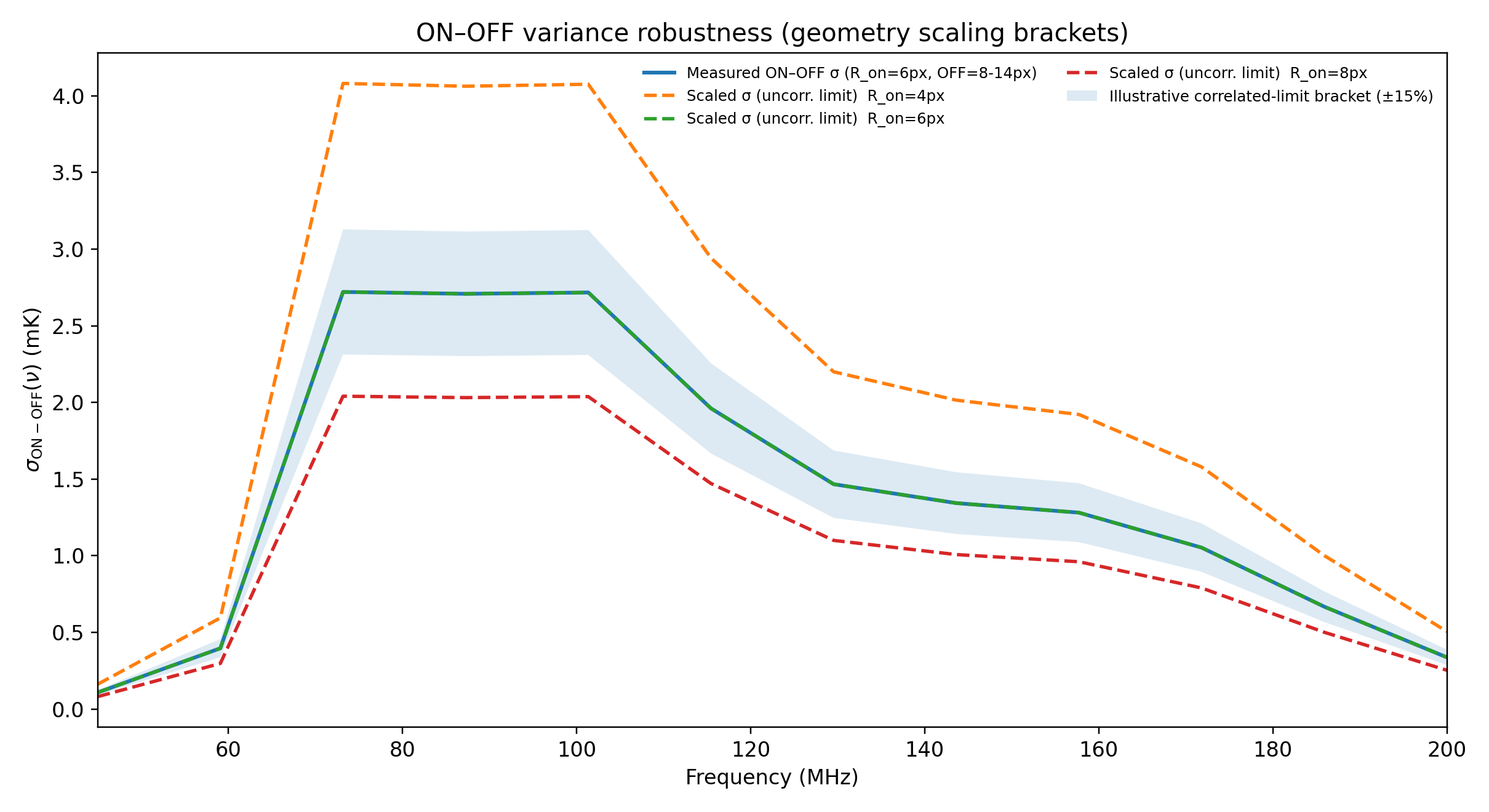}
\caption{Robustness of the coeval-derived ON--OFF variance layer to plausible choices of ON aperture and OFF annulus geometry. The plotted curves show a conservative bracket around the baseline $\sigma_{\rm astro}(\nu)$ by scaling the baseline variance to represent moderately smaller/larger effective ON and OFF regions. The overall frequency dependence and relative ordering are preserved, indicating that the variance budget used for separability is not driven by a fine-tuned geometric choice.}
\label{fig:geomrobust}
\end{figure}

\begin{figure}
\centering
\includegraphics[width=\columnwidth]{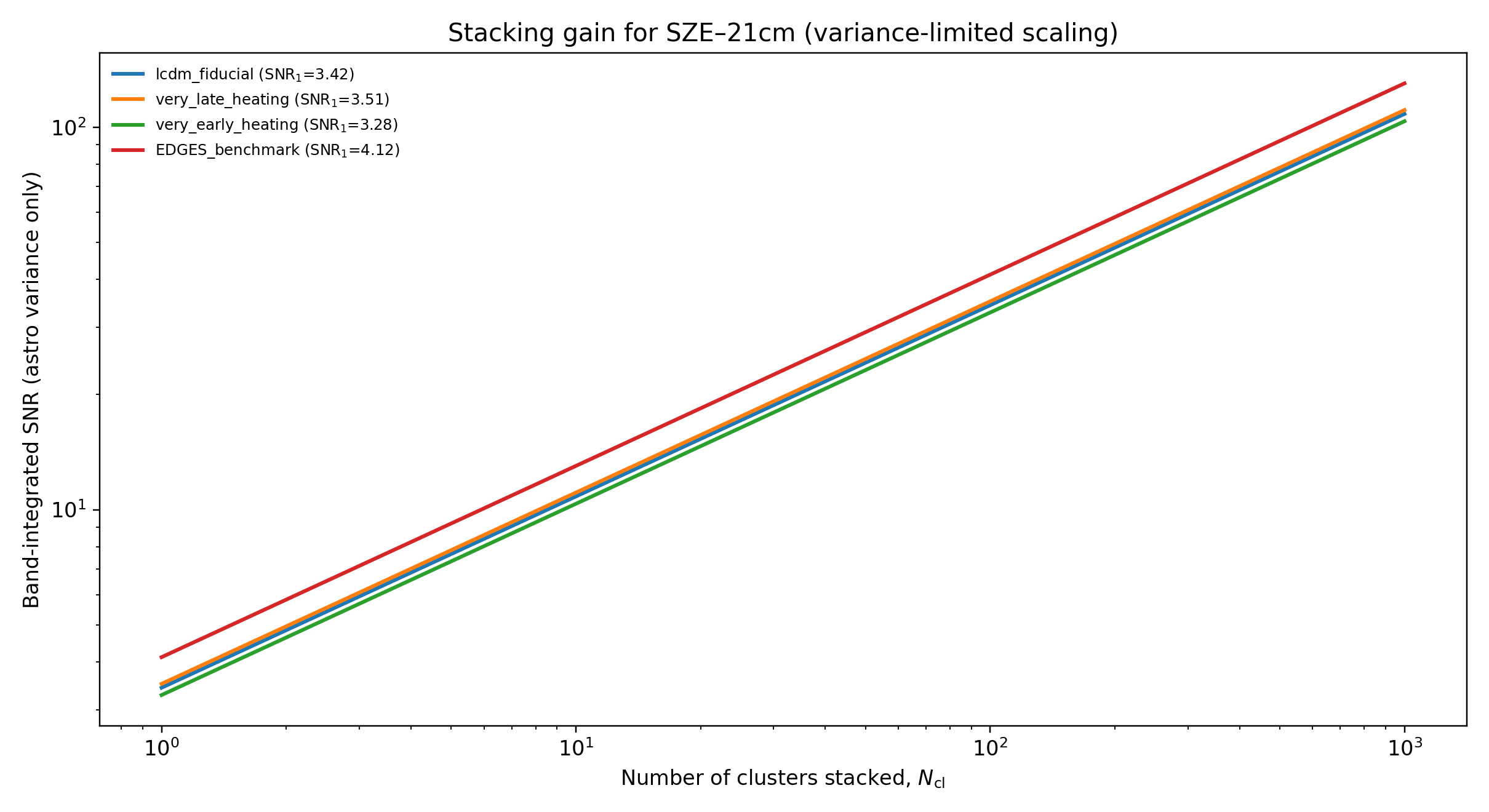}
\caption{Illustrative stacked signal-to-noise ratio scaling for SZE--21cm measurements as a function of the number of stacked clusters, $N_{\rm cl}$. SNR$_1$ denotes the band-integrated single-cluster SNR (cluster-by-cluster ratio of $\langle|\Delta T_{\rm SZE-21cm}|\rangle$ to $\sigma_{\rm astro}$, summed in quadrature across the 45--200\,MHz grid). Independent cluster sightlines are assumed, so the effective astrophysical variance scales as $1/\sqrt{N_{\rm cl}}$. This curve uses only $\sigma_{\rm astro}$ in the denominator and is therefore an optimistic, instrument-free upper bound; including thermal noise and foreground residuals would raise the required $N_{\rm cl}$ for a fixed SNR.}
\label{fig:stackingsnr}
\end{figure}

\begin{figure}
\centering
\includegraphics[width=\columnwidth]{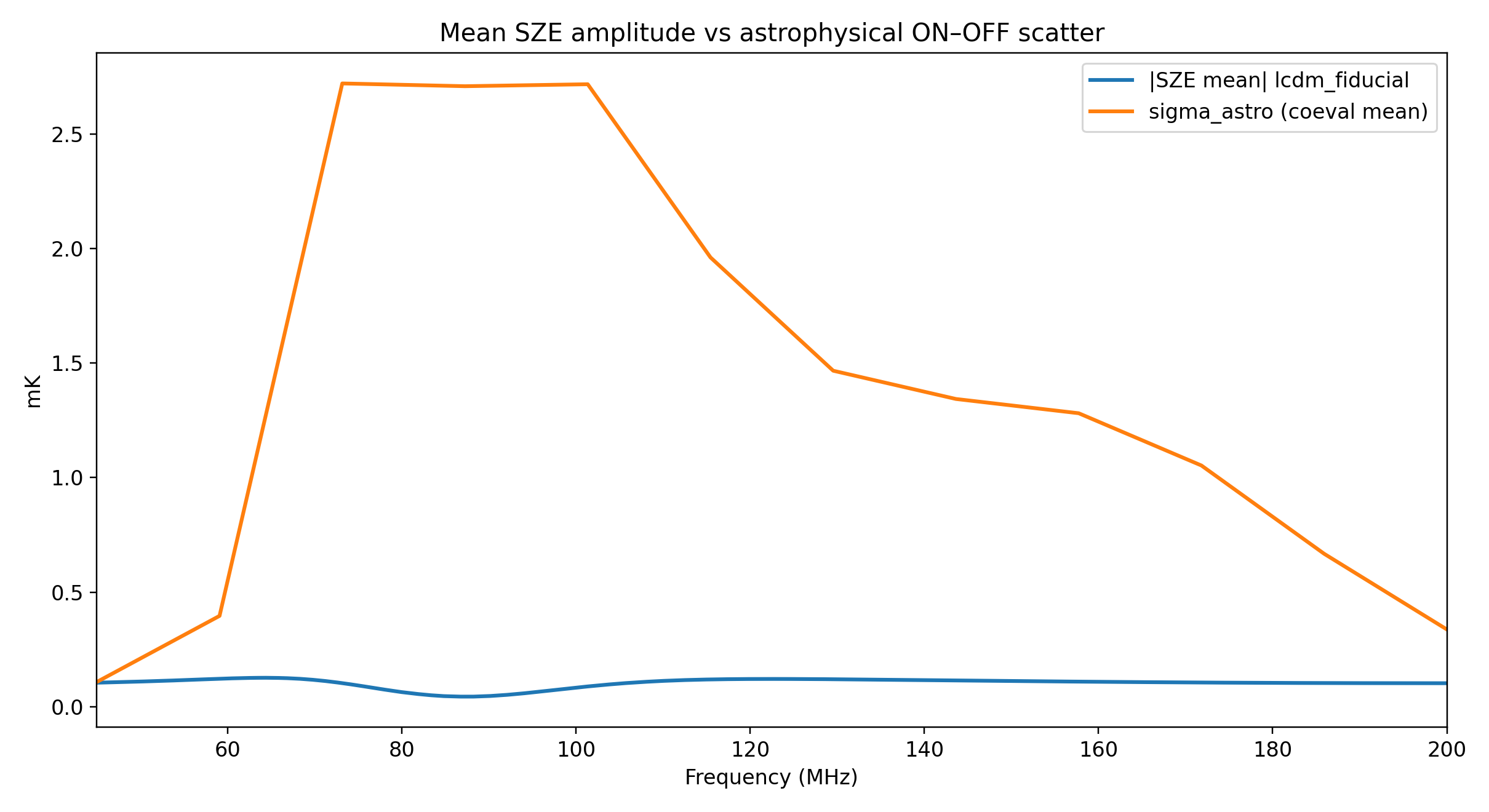}
\caption{Absolute mean SZE--21cm amplitude $|\langle\Delta T_{\rm SZE-21cm}\rangle|$ for the fiducial model compared with the coeval-derived variance layer $\sigma_{\rm astro}(\nu)$. Over much of the band, $\sigma_{\rm astro}$ exceeds the mean SZE amplitude, motivating stacking and conservative variance-aware discrimination metrics. (The plotted figure file labels these as ``$|$SZE mean$|$ lcdm\_fiducial'' and ``sigma\_astro (coeval mean)'' respectively.)}
\label{fig:meansigastro}
\end{figure}

Figure~\ref{fig:coevalsigma} shows the frequency-dependent ON--OFF scatter $\sigma_{\rm ON-OFF}(\nu)$ derived from coeval cubes (averaged over realisations as stored). We present the variance products in the Results section because $\sigma_{\rm astro}(\nu)$ is an explicit ingredient of the discrimination metric: it enters the band-integrated pairwise distance in Eq.~(\ref{eq:Dij}) and therefore directly determines variance-aware separability. The geometric robustness of $\sigma_{\rm astro}$ to ON/OFF aperture choice is shown in Fig.~\ref{fig:geomrobust}, the variance-limited stacking gain in Fig.~\ref{fig:stackingsnr}, and the comparison against the mean SZE--21cm amplitude in Fig.~\ref{fig:meansigastro}; together these motivate the conservative inclusion of $\sigma_{\rm astro}$ in the pairwise distance metric and point to stacking and optimised spatial filtering for robust detectability.

Figure~\ref{fig:coevalsigma} quantifies the intrinsic astrophysical variance expected when comparing an ON region centred on a cluster to an OFF annulus in nearby blank sky. The frequency dependence reflects the evolving morphology and amplitude of 21\,cm fluctuations across Cosmic Dawn. This statistic is observationally motivated: it is defined in terms of finite regions on the sky rather than box-wide variances, and it therefore provides a more realistic variance layer for differential SZE--21cm measurements than monopole seed scatter alone.

\section{Discussion}
\label{sec:discussion}

\subsection{Physical interpretation: what drives separability?}

The SZE--21cm is sensitive primarily to the curvature and structure of the incident low-frequency background rather than to a single-point measurement of $\delta T_b(\nu)$. Distinct astrophysical scenarios therefore map into separable SZE--21cm spectra when they shift the timing and shape of key Cosmic Dawn transitions:
\begin{itemize}
\item Heating timing changes (early vs late heating) strongly modify the depth and frequency location of the global absorption trough, and therefore leave the largest imprint in the SZE--21cm across 45--200\,MHz.
\item Star-formation efficiency modifies the overall emissivity and the transition timing, producing moderate but stacking-recoverable changes in the SZE--21cm signature space (see Sec.~\ref{sec:Dinterp} for the conversion to observational significance).
\item Suppressed sources partially mimic reduced emissivity, naturally leading to similarity with low-SFE scenarios, as captured by the small $D$ between these models.
\item X-ray hardness variations can be nearly degenerate in the SZE--21cm if the adopted parameter change does not substantially shift global-signal curvature across the band. This is not a statement that hardness is unimportant in general, but that the particular hardness contrast explored here does not move the observable sufficiently once astrophysical ON/OFF variance is considered.
\end{itemize}

\subsection{EDGES-like benchmarks and differential consistency tests}

Including an EDGES-like benchmark is informative because it provides a purely morphological stress test of how an anomalously deep global absorption feature would propagate into the SZE--21cm observable; we do not treat the EDGES-like curve as a physically motivated model within our parameter suite. In our calculations, the EDGES benchmark yields a deeper SZE--21cm minimum ($\Delta T_{\min}\approx -0.226$\,mK) than any physically motivated model in the current suite ($\Delta T_{\min}\approx -0.113$ to $-0.137$\,mK). The large pairwise distance from fiducial ($D\approx 1$) suggests that, in principle, differential observations of galaxy clusters could serve as an independent consistency check on the morphological class of the global-signal claim, especially when combined with stacking strategies and careful treatment of $\sigma_{\rm astro}(\nu)$. Unlike in earlier idealised demonstrations \cite{Takalana2020}, the flat-bottom morphology of the EDGES-like absorption feature is not preserved as a flat plateau in the corresponding SZE--21cm spectrum here. This is not an inconsistency in the formalism but reflects the intrinsic sensitivity of the SZE--21cm to spectral curvature rather than absolute signal depth: relativistic Comptonisation mixes photons across neighbouring spectral regions, so extended segments of near-zero curvature in the incident background are smoothed once scattering is taken into account. The SZE--21cm therefore responds primarily to the sharp edges and overall curvature of the EDGES-like feature, providing an independent morphological test of anomalous global signal shapes.

\subsection{Radio Sunyaev--Zel'dovich contamination from a cosmological excess radio background}
\label{sec:radioSZ}

The convolution operator that defines the SZE--21cm in Sec.~\ref{sec:kernel} is linear in the incident intensity. It therefore applies, with no modification, to any low-frequency background that fills the cluster's line of sight. If the ARCADE-2 excess radio monopole \cite{Fixsen2011} is cosmological in origin, this includes a smooth radio background of brightness temperature $T_R(\nu) = T_{R,0}(\nu/310\,{\rm MHz})^{-\gamma_R}$ with $T_{R,0}=24.1$\,K and $\gamma_R=2.59$. Its Comptonisation by the cluster produces the radio Sunyaev--Zel'dovich effect~\cite{HolderChluba2021,LeeChlubaHolder2022}.

\begin{figure}
\centering
\includegraphics[width=\columnwidth]{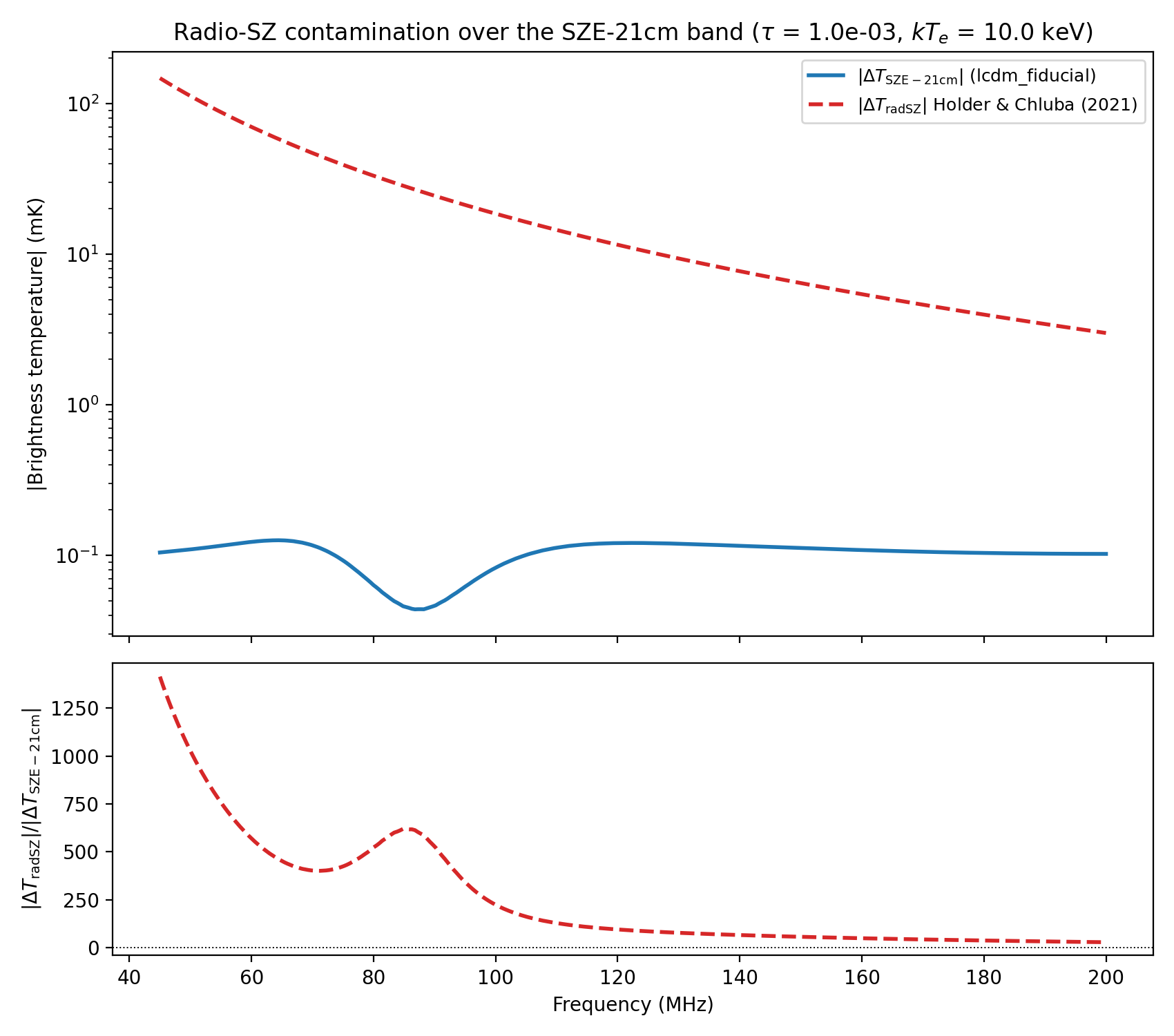}
\caption{Radio Sunyaev--Zel'dovich contamination over the SZE--21cm band, for a fiducial cluster with $\tau_0=10^{-3}$ and $kT_e=10$\,keV. \emph{Top:} $|\Delta T_{\rm SZE-21cm}|$ for the fiducial \textsc{21cmFAST} model (solid blue) and the Holder \& Chluba~\cite{HolderChluba2021} analytic radio-SZ form (dashed red). The radio SZ exceeds the SZE--21cm by one to three orders of magnitude across the band. \emph{Bottom:} contamination ratio $|\Delta T_{\rm radSZ}|/|\Delta T_{\rm SZE-21cm}|$ versus frequency.}
\label{fig:radioSZ}
\end{figure}

In the non-relativistic, low-frequency limit, Holder \& Chluba~\cite{HolderChluba2021} give a closed form in brightness temperature,
\begin{equation}
\frac{\Delta T_{\rm radSZ}(\nu)}{T_{\rm CMB}}
\simeq 2y\left[\left(\frac{\nu}{\nu_{\rm null}}\right)^{-\gamma_R}-1\right],
\label{eq:radioSZ_HC}
\end{equation}
with $\nu_{\rm null}\simeq 735$\,MHz under the cosmological-ERB hypothesis and $y=\tau_0\theta_e$ the cluster Compton parameter. For our representative cluster ($\tau_0=10^{-3}$, $kT_e=10$\,keV, $y\simeq 1.96\times 10^{-5}$) this gives $|\Delta T_{\rm radSZ}|\simeq 150$\,mK at $45$\,MHz, $\simeq 19$\,mK at $100$\,MHz, and $\simeq 3$\,mK at $200$\,MHz. The corresponding ratio to the SZE--21cm amplitude in our fiducial model has a band-median value of $\sim 90$ and reaches a maximum of $\sim 1.4\times 10^{3}$ at the low-frequency end of the band. Figure~\ref{fig:radioSZ} shows the full spectrum across $45$--$200$\,MHz.

Two consequences follow. First, the radio SZ does not cancel in the ON--OFF differential strategy adopted in this work: the smooth ERB monopole cancels, but its Comptonisation by the cluster appears only in the ON aperture, by the same logic that justifies the SZE--21cm differential. Second, if the ERB is cosmological at the ARCADE-2 amplitude, the radio SZ exceeds the SZE--21cm by one to three orders of magnitude across the band, and a future SZE--21cm detection from cluster stacking would have to be combined with an independent constraint on the ERB amplitude, for example, a higher-frequency radio-SZ measurement near $\nu_{\rm null}$ as suggested by~\cite{HolderChluba2021,LeeChlubaHolder2022}, or a CMB SZE or X-ray determination of the cluster's Compton-$y$ that allows subtraction of the radio component.

A cosmological ERB at this level also alters the 21\,cm signal $\delta T_b$ itself, by modifying the radiative background that sets the spin temperature and the contrast against the photon field \cite{FengHolder2018}. Our \textsc{21cmFAST} runs adopt the standard CMB-only radio background and therefore do not include this effect; an ERB-amplified $\delta T_b$ would be deeper than the global signals shown in Fig.~\ref{fig:globaloverlay}, and would propagate through the SZE--21cm operator independently of the radio-SZ contamination quantified above. Recent forward-modelling of the EDGES sky-temperature data with a physical Pop~III radio-background model, rather than the original flattened-Gaussian template, disfavours the presence of a non-standard $\sim$500\,mK absorption feature in the data \cite{Cang2025}, reinforcing our treatment of the EDGES-like template as a morphological stress test rather than a physical model.

\subsection{Interpretation of the pairwise distance and what \texorpdfstring{$D\sim 0.15$--$0.28$}{D approx 0.15-0.28} means in practice}
\label{sec:Dinterp}

The pairwise distances reported in Section~\ref{sec:results} are deliberately constructed to be intrinsic, instrument-free separability indices. By construction the denominator of Eq.~(\ref{eq:Dij}) contains only (i) the seed-to-seed scatter $\sigma_{\rm seed}$, which is small in the monopole pipeline, and (ii) the astrophysical ON--OFF variance $\sigma_{\rm astro}$ from coeval cubes. We have not added a thermal-noise term, because doing so requires a specific instrument and integration time and would tie the metric to one assumed observing strategy. The discussion below also assumes that the radio SZ contribution from a possible cosmological excess radio background, quantified in Sec.~\ref{sec:radioSZ}, has been independently constrained and subtracted via the strategies discussed there (e.g.\ a higher-frequency radio-SZ measurement near $\nu_{\rm null}\simeq 735$\,MHz, or a CMB SZE / X-ray determination of the cluster's Compton-$y$); a residual radio-SZ contamination would enter as an additional contribution to the variance budget and require correspondingly larger cluster samples for the same $D_{\rm obs}$.

A direct consequence is that $D$ here is an upper bound on what any real experiment with finite thermal noise will achieve. To translate the values into observational language, note that for a stacked observation of $N_{\rm cl}$ clusters with independent ON/OFF sightlines and per-channel thermal noise $\sigma_{\rm therm}$, the effective per-channel variance is
\begin{equation}
\sigma_{\rm tot}^2(\nu)\;\simeq\;\frac{\sigma_{\rm astro}^2(\nu)+\sigma_{\rm therm}^2(\nu)}{N_{\rm cl}},
\label{eq:sigmatot}
\end{equation}
and the corresponding observational separability scales as $D_{\rm obs}\simeq D\sqrt{N_{\rm cl}\,\sigma_{\rm astro}^2/(\sigma_{\rm astro}^2+\sigma_{\rm therm}^2)}$. With $\langle\sigma_{\rm astro}\rangle\sim 1.4$\,mK over our band, the heating-timing pair (\texttt{very\_late\_heating} versus \texttt{lcdm\_fiducial}, $D\simeq 0.28$) reaches $D_{\rm obs}\simeq 3$ --- a $3\sigma$-equivalent broadband separation --- for $N_{\rm cl}\simeq 100$ in the variance-limited regime, growing to $\simeq 9$ for $N_{\rm cl}=1000$. The star-formation-efficiency pair ($D\simeq 0.15$) requires roughly four times more clusters to reach the same broadband significance, while the soft/hard X-ray contrast in the present parametrisation ($D\sim 0.005$) is not separable on these grounds without supplementary information. We therefore use ``measurable'' in this section only in this stacking-limited, idealised sense, and do not claim that single-cluster SZE--21cm measurements can resolve modest SFE variations on their own.

A full instrument-level detectability forecast requires sensitivity curves and end-to-end assumptions about beam and $uv$ response, calibration, foreground mitigation, band-averaging, and integration time, and lies outside the scope of this paper. The variance-aware framework developed here is the natural input to such a forecast: $\sigma_{\rm astro}(\nu)$ from coeval ON/OFF statistics ranges over $\mathcal{O}(0.1$--$3)$\,mK across the band and exceeds the mean SZE--21cm amplitude ($\mathcal{O}(0.1)$\,mK) for single clusters over most of 45--200\,MHz, so stacking and matched spatial filtering will be essential. The pairwise heatmap framework already encodes separability under an explicit variance budget, and a follow-up forecast simply adds a thermal-noise term in the denominator of Eq.~(\ref{eq:Dij}) for a chosen instrument and integration time.

\subsection{Non-relativistic kernel comparison}
\label{sec:nonrel}

\begin{figure}
\centering
\includegraphics[width=\columnwidth]{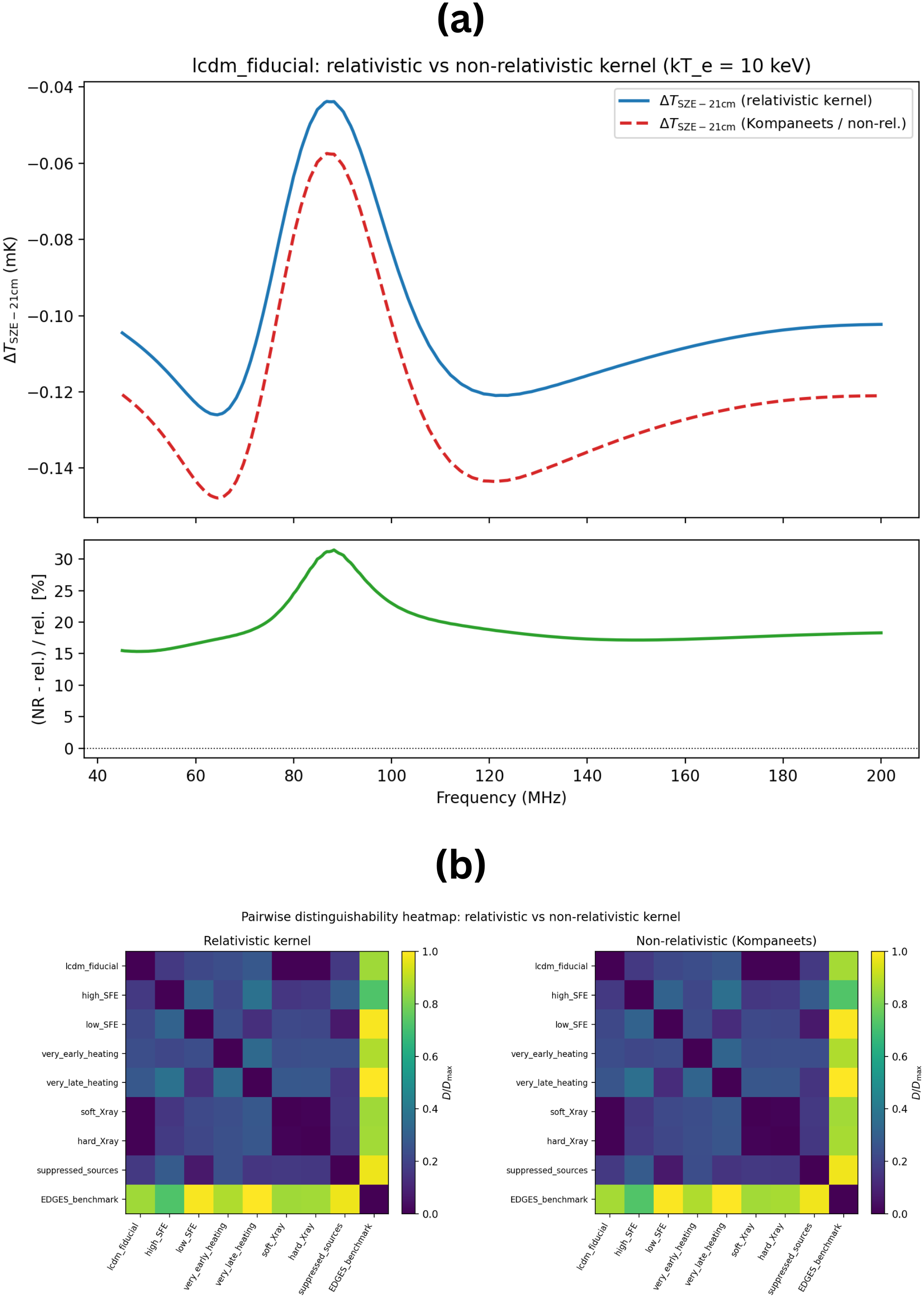}
\caption{Non-relativistic kernel comparison. \emph{(a)} SZE--21cm spectrum for the fiducial model computed with the full relativistic kernel $P(s)$ (solid blue) and with a Gaussian kernel matched to the leading thermal moments, $P_{\rm NR}(s)=\mathcal{N}(\mu=3\theta_e,\sigma^2=2\theta_e-\mu^2)$ (dashed red), together with the fractional difference. \emph{(b)} pairwise distinguishability heatmap $D_{ij}/D_{\max}$ recomputed under $P_{\rm NR}$ alongside the relativistic version. The model ranking is preserved; entries shift by at most $\sim$15\%.}
\label{fig:nonrel}
\end{figure}

To check how much of the SZE--21cm spectrum is set by the relativistic structure of the kernel rather than by its first two moments, we recompute all eight \textsc{21cmFAST} models with a Gaussian kernel matched to the leading thermal moments,
$P_{\rm NR}(s)=\mathcal{N}(s;\mu=3\theta_e,\sigma^2=2\theta_e-\mu^2)$,
in place of $P(s)$ in Eq.~(\ref{eq:I1}). Figure~\ref{fig:nonrel} shows the comparison for the fiducial model together with the pairwise distinguishability heatmap recomputed under $P_{\rm NR}$. The spectral shape is preserved across the band, the minimum frequency shifts by less than $2$\,MHz, and the minimum depth shifts by $15$--$18\%$ across the eight \textsc{21cmFAST} models at $kT_e=10$\,keV --- the non-relativistic kernel systematically overestimates the depth by this amount, consistent with the pattern reported by Colafrancesco et al.~\cite{Colafrancesco2016}. The frequency-resolved difference $(\mathrm{NR}-\mathrm{rel})/\mathrm{rel}$ is not uniform across the band: it is $\sim$15\% over much of the range and rises to $\sim$30\% at the frequencies of maximum spectral curvature (lower panel of Fig.~\ref{fig:nonrel}), since the non-relativistic kernel reproduces the incident curvature least accurately where that curvature is largest. The pairwise $D_{ij}$ ranking is preserved exactly: under both kernels the six most-separated models from fiducial are, in order, \texttt{EDGES\_benchmark}, \texttt{very\_late\_heating}, \texttt{very\_early\_heating}, \texttt{low\_SFE}, \texttt{high\_SFE}, and \texttt{suppressed\_sources}. The relative entries of $D_{ij}/D_{\max}$ shift by at most $\sim$15\%. At $kT_e=10$\,keV the model ranking and overall separability indices are therefore robust to the choice of kernel, but the non-relativistic kernel systematically misestimates the minimum depth by $\sim$15--18\%, and this error grows with cluster temperature. We retain the relativistic kernel as the working choice for the remainder of the analysis and recommend its use in any extension to hotter or more heterogeneous cluster samples.

\subsection{Positioning among Cosmic Dawn probes}

The results presented here demonstrate that the SZE--21cm is not a replacement for established 21\,cm probes such as power-spectrum measurements, but rather a complementary, differential observable that is sensitive to the morphology of the global background rather than to its absolute monopole. For physically motivated Cosmic Dawn scenarios, the SZE--21cm responds most strongly to changes in heating timing and overall emissivity, while remaining comparatively insensitive to modest variations in X-ray spectral hardness. This behaviour is consistent with the convolutional nature of the Comptonisation process, which maps broadband curvature and transition timing in the incident spectrum into cluster-localised distortions.

Importantly, when an astrophysical ON--OFF variance layer derived from coeval cubes is included, the expected variance can exceed the mean SZE--21cm amplitude for single clusters over much of the band. This indicates that robust detection and model discrimination will likely require stacking over multiple clusters and/or optimised spatial filtering. Nevertheless, the variance-aware pairwise distinguishability analysis shows that several classes of Cosmic Dawn models remain separable in principle. In this sense, the SZE--21cm provides a physically independent consistency test on early-Universe heating histories that is naturally compatible with interferometric observations, complementing global-signal and power-spectrum approaches rather than duplicating them.

\section{Conclusions}
\label{sec:conclusions}

We have presented a variance-aware analysis of how well different Cosmic Dawn scenarios can be separated using the SZE--21cm over 45--200\,MHz. Across a suite of physically motivated Cosmic Dawn scenarios, the predicted SZE--21cm spectra exhibit meaningful shifts in characteristic frequencies and changes in minimum depth at the $\sim10$--$20$ per cent level, with heating-timing variations producing the most distinct broadband signatures. The SZE--21cm minimum does not, in general, coincide with the minimum of the incident global 21\,cm signal, emphasising that the cluster distortion responds to broader spectral curvature through relativistic Comptonisation rather than to a single-point feature.

A key result is that, while the seed-to-seed scatter in the monopole-based pipeline is extremely small, a second variance layer derived from coeval ON/OFF statistics is substantially larger and must be included for conservative separability (and, ultimately, detectability) assessments. Incorporating this variance layer into the band-integrated separability index yields a pairwise distinguishability heatmap that provides a compact and transparent summary of which physical scenarios are intrinsically distinguishable in the SZE--21cm observable itself. Within the parameter choices explored here, heating timing and the EDGES-like benchmark are most distinct from the fiducial model, whereas the specific soft/hard X-ray variants remain nearly degenerate once the astrophysical ON/OFF variance is included. The framework developed here, combining coeval-derived variance products with full-spectrum pairwise diagnostics, provides a practical basis for designing and interpreting stacked cluster SZE--21cm measurements as observing strategies and instrumental sensitivities are refined.

\acknowledgments
The author acknowledges support from the International Astronomical Union Office of Astronomy for Development (IAU--OAD), the South African Astronomical Observatory (SAAO), and Stellenbosch University.

\end{document}